\documentclass[twocolumn,trackchanges]{aastex701}
\usepackage{acro}
\acsetup{patch/longtable=false} 
\usepackage{graphicx}	
\usepackage{amsmath, amssymb, graphicx, color, units, xspace}
\usepackage{hyperref}
\usepackage{fontawesome5}
\usepackage{xcolor}
\usepackage[T1]{fontenc}
\usepackage{subcaption}

\graphicspath{ {./} }

\newcommand{\mone}{\ensuremath{M_{1}}\xspace}
\newcommand{\mtwo}{\ensuremath{M_{2}}\xspace}
\newcommand{\q}{\ensuremath{q}\xspace}

\newcommand{\kms}{\ensuremath{\,\rm{km}\,\rm{s}^{-1}}\xspace}
\newcommand{\Msun}{\ensuremath{\,\rm{M}_{\odot}}\xspace}

\received{May 20, 2026}
\revised{October 11, 2026}
\accepted{October 12, 2026}

\DeclareAcronym{UFD}{
  short = UFD ,
  long  = ultra-faint dwarf
}
\DeclareAcronym{BHNS}{
  short = BHNS ,
  long  = black hole--neutron star
}

\DeclareAcronym{BNS}{
  short = BNS ,
  long  = binary neutron star
}

\DeclareAcronym{BBH}{
  short = BBH ,
  long  = binary black hole
}

\DeclareAcronym{NS}{
  short = NS ,
  long  = neutron star
}

\DeclareAcronym{BH}{
  short = BH ,
  long  = black hole
}

\DeclareAcronym{LVK}{
  short = LVK ,
  long  = LIGO-Virgo-KAGRA
}

\DeclareAcronym{GW}{
  short = GW ,
  long  = gravitational wave
}

\DeclareAcronym{CE}{
  short = CE ,
  long  = common-envelope
}

\DeclareAcronym{SMT}{
  short = SMT ,
  long  = stable mass transfer
}

\DeclareAcronym{DCO}{
  short = DCO ,
  long  = double-compact object
}

\DeclareAcronym{ZAMS}{
  short = ZAMS ,
  long  = zero-age main sequence
}

\DeclareAcronym{MRR}{
  short = MRR ,
  long  = mass ratio reversal
}

\begin{document}

\title{Massquerade: Impacts of Mass Ratio Reversals on Binary Black Hole Merger Rates and Mass Distributions}

\author[orcid=0000-0002-1732-8040,sname='North America']{Tyler B. Smith}
\affiliation{Department of Physics and Astronomy
University of California, Irvine, CA 92697, USA}
\email[show]{tylerbs@uci.edu}  

\author[orcid=0000-0002-4421-4962,sname='North America']{Floor Broekgaarden}
\affiliation{Department of Astronomy and Astrophysics
University of California, San Diego, CA 92093, USA}
\email[]{fbroekgaarden@ucsd.edu}

\author[orcid=0000-0003-1241-7615,sname='North America']{Sasha Levina}
\affiliation{Department of Astronomy and Astrophysics
University of California, San Diego, CA 92093, USA}
\email[]{slevina@ucsd.edu}

\author[orcid=0000-0001-9583-4339,sname='North America']{Amedeo Romagnolo}
\affiliation{Universität Heidelberg, Zentrum für Astronomie (ZAH), Institut für Theoretische Astrophysik, Albert Ueberle Str. 2, 69120, Heidelberg, Germany}
\affiliation{Dipartimento di Fisica e Astronomia Galileo Galilei, Università di Padova, Vicolo dell’Osservatorio 3, I–35122 Padova, Italy}
\email[]{amedeo.romagnolo@uni-heidelberg.de}

\author[orcid=0009-0007-7586-5403,sname='North America']{Manasvini Komandur}
\affiliation{Department of Astronomy and Astrophysics
University of California, San Diego, CA 92093, USA}
\email[]{mkomandur@ucsd.edu}

\author[orcid=0009-0001-1988-8383,sname='North America']{Melanie Santiago}
\affiliation{Department of Astronomy and Astrophysics
University of California, San Diego, CA 92093, USA}
\email[]{m6santiago@ucsd.edu}

\author[orcid=0000-0003-4474-6528,sname='North America']{Kyle A. Rocha}
\affiliation{Department of Astronomy and Astrophysics
University of California, San Diego, CA 92093, USA}
\email[]{kyrocha@ucsd.edu}

\begin{abstract}
Mass ratio reversal (MRR), in which the initially less massive star in a binary ultimately forms the more massive compact object, is a well-established outcome of interacting binary evolution. We investigate its role in shaping the binary black hole (BBH) merger rate and mass distribution inferred by LIGO--Virgo--KAGRA, using population synthesis frameworks \texttt{COMPAS} and \texttt{SEVN}. The observational imprint of MRR differs qualitatively: in \texttt{COMPAS}, MRR systems dominate the merger rate density at high primary masses (\mone $\gtrsim$ 20 M$_\odot$), secondary masses (\mtwo $\gtrsim$ 12 M$_\odot$), and mass ratios ($q>0.6$), whereas in \texttt{SEVN} they remain subdominant across the BBH mass distribution. This implies that the initially less massive star can \textit{massquerade} as the observed primary black hole, such that the primary-mass distribution is not a direct tracer of the initially more massive stars, but instead a superposition of physically distinct evolutionary populations. We identify three evolutionary pathways leading to MRR: \textit{core-growth}, where stable mass transfer increases the helium-core of the secondary; \textit{PPISN-shrinking}, where pulsational pair-instability episodes reduce the primary remnant mass; and \textit{asymmetric-CCSN}, where differential supernova mass loss drives the reversal. When weighted by the local merger-rate density, the core-growth channel dominates. MRR systems predominantly originate from massive ($\gtrsim$ 50 M$_\odot$) low-metallicity progenitors ($ \lesssim$ 0.1 $Z_\odot$). MRR is a physically distinct, potentially observable feature of isolated binary evolution that must be accounted for to robustly connect future gravitational-wave observations to massive binary evolution and compact-object formation.

\end{abstract}


\section{Introduction} 

The LIGO--Virgo--KAGRA (LVK) collaboration has detected over 250 confirmed gravitational-wave events through the second part of observing run 4 (O4b), primarily from binary black hole (BBH) mergers \citep{GWTC1,Abbott2020c,GWTC3,GWTC4pop,gwtc5-population}. These observations are commonly summarized in terms of the total merger rate density (MRD), $\mathcal{R}(z)$, as well as differential merger rates with respect to intrinsic binary parameters, such as the component BH masses. 
In particular, LVK population analyses characterize the BBH population using the primary- and secondary-mass distributions, $d\mathcal{R}/dM_1$ and $d\mathcal{R}/dM_2$, where the primary (secondary) black hole is defined as the more (less) massive component of the system. 

From the perspective of binary evolution, however, the two compact objects often experience systematically different evolutionary histories: the initially more massive star typically evolves first and acts as the donor during the first mass-transfer phase, while the initially less massive star may accrete mass and angular momentum before collapsing \citep{Kippenhahn:1967,Tauris:2006,vandenHeuvel:2017,Renzo:2021}. As a result, one may expect the observed $M_1$ and $M_2$ distributions to approximately trace two distinct evolutionary populations of donors and accretor stars. 

This mapping can break down in the presence of \textit{mass ratio reversal} (MRR), where the initially less massive star ultimately forms the more massive black hole. This outcome is not unusual, but rather a well-established outcome of short-period binary evolution. In particular, systems such as Algol binaries exhibit clear observational evidence of MRR \citep[e.g.][]{Crawford:1955,Morton:1960,Smak:1962,plavec:1968,paczynski:1971,Budding:1986,Harries:2003,deMink:2007SMC,Sen:2022,Sen:2023,Sen:2026}. This phenomenon has also been observed in compact-object studies, e.g. in neutron star--white dwarf (NS--WD) binaries \citep{Tauris:2000,toonen18nov}, BH--NS binaries \citep{Sipior:2004,Broekgaarden:2021imbI}, and BBH studies \citep{Gerosa:2013,Gerosa:2018prd,Olejak:2021iux,Broekgaarden:2022,ZevinBavera:2022,Olejak:2024,Banerjee:2024,Smith:2026}, demonstrating that MRR is a generic consequence of binary evolution. Recently, MRR has been studied in the context of gravitational-wave observations where characteristic mass-spin correlations have been used to constrain the fraction of LVK sources which arise from MRR \citep{mould22dec}. The majority of BBH studies focus on MRR as related to the spins of BBH systems. For instance, \citet{Broekgaarden:2022} found that MRR can occur in  $\gtrsim 70\%$ of observable LVK sources and between 11--82$\%$ of the astrophysical population, depending on stellar, binary, and star formation prescriptions used, with MRR systems preferentially populating the primary-mass spectrum above $\sim 20  \Msun$, and the second-born BH acquiring non-negligible spin in up to $\sim 25\%$ of BBH systems. The latter finding is consistent with expectations that the initially more massive star produces a weakly spinning remnant, while the secondary star may be tidally spun up prior to collapse \citep{kushnir16oct,hotokezakajun17,Qin:2018vaa,fuller19aug,bavera22sep,belczynski20apr}. 

Building on these results, we investigate how MRR shapes the BBH merger-rate and mass distributions predicted by population-synthesis models. We quantify the MRR contribution to the BBH merger rate as a function of redshift, primary mass, secondary mass, and mass ratio, and assess the implications for interpreting BBH observations.
We further examine the progenitor properties and formation channels of MRR BBH systems, identifying the physical mechanisms driving MRR and their BBH mass distribution signatures. Finally, we compare predictions from two fundamentally different population-synthesis frameworks: \texttt{COMPAS} and \texttt{SEVN}, to investigate how  assumptions in stellar and binary evolution shape the prevalence and observational imprint of MRR BBH systems.

\section{Methodology}
\label{sec:methods}

\subsection{Population Synthesis Models}
\label{subsec:pop-synth}

In order to determine the MRR contribution to the astrophysical merger rate, we first construct a population of merging BBHs using two distinct population synthesis frameworks, \texttt{COMPAS} and \texttt{SEVN}, which employ contrasting approaches to stellar evolution (analytic fitting formulae versus interpolating tracks \textit{on-the-fly}, respectively).
Specifically, we include the fiducial and single-parameter variation \texttt{COMPAS} models from \cite{Broekgaarden:2022imbII}, in which MRR has been studied in \citet{Broekgaarden:2022} finding that MRR comprises $11$--$82\%$ of the astrophysical population depending on the model choice. We compare this to the results from \texttt{SEVN}, where MRR systems were previously found to comprise $\sim20\%$ of the astrophysical population \citep{Smith:2026}. These studies indicate that there can be upwards of $60\%$ difference in their calculated MRR contributions. 

We note that the two codes will have inherent differences as they approach stellar evolution in contrasting ways, i.e. \texttt{COMPAS} employs analytic fitting formulae from \citet{Hurley:2000} based on the tracks from \citet{Pols:1998} (with masses up to 50 \Msun at 7 distinct metallicities) vs interpolating these quantities \textit{on-the-fly} from PARSEC tracks \citep{Bressan2012,Tang2014,Chen2015,Marigo2017,Nguyen2022} (with masses up to 600 \Msun at 15 distinct metallicities) in \texttt{SEVN}. We perform a systematic comparison with a consistent cosmic integration, i.e. star-formation history. We extend the two studies mentioned above by including the primary-mass, secondary-mass, and mass-ratio distributions with a full MRR contribution decomposition.

\subsubsection{COMPAS simulations}
\label{subsec:compas}
\texttt{COMPAS} \citep{Stevenson:2017tfq,Barrett:2017fcw,Vigna-Gomez:2018dza,Broekgaarden:2019qnw,Neijssel_2019} utilizes parameterized models of binary evolution from \citet{Hurley:2002rf}. Specifically, we use the publicly available models from \citet{Broekgaarden:2022imbII}, which include 20 single parameter variation on massive (binary) star evolution and are publicly available at  \citet[][]{Broekgaarden:2021-zenodo-BHBH}. In particular, we use the fiducial dataset and variations (B,C,D,G,I,J,L,P,Q,S, and T) from \citet{Broekgaarden:2022imbII}, chosen for their effects on mass transfer and binary survivability. The initial parameters are sampled according to \citet{Sana2012} within the following ranges: $\log_{10}(Z/Z_\odot) \in [-2.3,  -0.38]$, $M_{1,\mathrm{ZAMS}}/\Msun \in [5,150]$, mass ratio $q = M_2/M_1 \in [q_{\rm min},1]$ ($q_{\rm min} = \rm max(2.2/M_1 , 0.1)$), and orbital separation $a/\rm AU \in [0.01,1000]$.

\subsubsection{SEVN simulations }
\label{subsec:sevn}
We compare these results with \texttt{SEVN}, which calculates stellar properties via the interpolation of precomputed evolutionary tracks \citep{Spera2015,Spera2017,Spera2019,Mapelli2020,Iorio:2023sevn}. We use version \texttt{SEVN v2.13.0} with the default parameters adopted in the {\tt PARSEC} tracks with overshooting parameter $\lambda=0.5$ \citep{Bressan2012,Tang2014,Chen2015,Marigo2017,Nguyen2022}. The initial parameters were sampled following \citet{Smith:2026} based on \citet{Sana2012} in the following ranges: $\log_{10}(Z/Z_\odot) \in [-4.0,  0.18]$, $M_{1,\mathrm{ZAMS}}/\Msun \in [3,200]$, $q \in [0.05,1]$, and $a/\rm AU \in [0.07,670]$.

\subsubsection{Fiducial binary and single-star physics assumptions}
\label{subsec:defaults}

Beyond the initial-condition sampling described above, the treatment of common-envelope evolution, mass transfer, and supernova physics adopted in each code shapes which binaries survive to merge and the remnant masses of the resulting black holes. Below we summarize the fiducial binary- and single-star physics assumptions in \texttt{COMPAS} and \texttt{SEVN}. 

In both \texttt{COMPAS} and \texttt{SEVN} the standard $\alpha$--$\lambda$ formalism is used to parameterize the common-envelope phase. The fiducial common-envelope efficiency is $\alpha_{\rm CE}=1.0$ in \texttt{COMPAS} and $\alpha_{\rm CE}=3.0$ in \texttt{SEVN}, while the binding-energy parameter $\lambda_{\rm CE}$ follows \citet{Xu2010,Xu2010b} in \texttt{COMPAS} and \citet{Claeys2014} in \texttt{SEVN}.

Mass transfer is Eddington-limited for compact-object accretors in both codes. For non-degenerate (stellar) accretors, which includes the main-sequence secondaries that grow their cores in the core-growth channel, the treatments differ. In \texttt{SEVN}, the fraction of the mass lost by the donor that is accreted by the secondary, the mass-transfer efficiency, is fixed at $\beta=0.5$. In the fiducial \texttt{COMPAS} model, accretion onto stellar accretors is instead thermal-timescale limited: the maximum accretion rate is set by the accretor’s thermal timescale and depends on its stellar type, while the mass-transfer rate is set by the donor’s thermal timescale. We refer the reader to \citet{compas2022} for full details.

The two codes also adopt different mass-transfer stability criteria. In \texttt{COMPAS}, stability is determined by comparing the response of the donor's radius to mass loss, $\zeta_\star \equiv d\ln R_\star/d\ln M$, to the response of the Roche-lobe radius to mass transfer, $\zeta_{\rm RL}\equiv d\ln R_{\rm RL}/d\ln M$ \citep[][and references therein]{compas2022}. In \texttt{SEVN}, the mass ratio of donor to accretor, $q_{\rm MT}\equiv M_{\rm d}/M_{\rm a}$, is directly compared to a critical mass ratio $q_{\rm crit}$; mass transfer is considered dynamically unstable whenever $q_{\rm MT}>q_{\rm crit}$. The specific values of $\zeta_\star$ and $q_{\rm crit}$ adopted for each stellar type are given in \citet{compas2022} and \citet{Iorio:2023sevn}, respectively.

Stellar winds in \texttt{SEVN} are accounted for directly in the precomputed stellar tracks; we refer the reader to \citet{Spera2015} for their effect on the tracks, while wind mass transfer between binary components follows \citet{Hurley:2002rf}. In \texttt{COMPAS}, winds are implemented following \citet{Belczynski2010,Belczynski2010b}.

Both codes adopt the delayed supernova engine of \citet{Fryer:2012} to determine remnant masses from the pre-supernova core structure. At the highest masses, pair-instability supernovae (PISN) and pulsational PISN (PPISN) are implemented in \texttt{COMPAS} following \citet{Marchant:2019}, and in \texttt{SEVN} following \citet{Mapelli2020}, who adopt a fitting formula based on \citet{Woosley2017}. In both codes, the natal supernova kick is drawn from a Maxwellian distribution with 1D dispersion $\sigma^{\rm 1D}_{\rm rms}=265\,\kms$. We note that \citet{Disberg:2025} show the original calculation of this value omitted a Jacobian, so that it is overestimated, and the corrected value is $\sim 220\,\kms$. We test substantially lower kick dispersions in \texttt{COMPAS} (e.g. $100\,\kms$ and $30\,\kms$) and find that this does not qualitatively change the role of MRR, and thus our conclusions.

Several of these assumptions, e.g. $\alpha_{\rm CE}$, the mass-transfer efficiency, and the natal-kick dispersion, are varied individually for \texttt{COMPAS} in Section~\ref{sec:results-impact-uncertainties-compas}; the values above are the fiducial choices adopted throughout the rest of this work, and this is not an exhaustive account of every prescription in either code. The complete set of fiducial parameters is given in Table~1 of \citet{Broekgaarden:2022imbII} for \texttt{COMPAS} and throughout  \citet{Iorio:2023sevn} for \texttt{SEVN}.

\subsection{Reanalyzing the Simulations Under Consistent Star Formation History Modeling}
In order to mitigate discrepancies due to the chosen cosmic integration, i.e. the  assumptions for the metallicity-dependent star formation history to calculate cosmological BBH merger rates, we use the \texttt{SSPC} \citep{Hendriks:2023yrw} code and recalculate the BBH merger properties across cosmic time for both the \texttt{SEVN} and \texttt{COMPAS} simulations under similar parameter settings, an exploration of single-parameter variations is provided in the results section. 
This allows for a standardized cosmic integration based on ingredients such as the star formation rate and metallicity-mass relation. In particular, in population synthesis studies the BBH merger rate can be calculated by a convolution of the BBH formation efficiency and a cosmic star formation history following

\begin{equation}
\begin{split}
\mathcal{R}_{\mathrm{BBH}}(z_{\mathrm{merge}},\, \zeta)
    = \int_{Z_{\min}}^{Z_{\max}} dZ
      \int_{0}^{t^{\ast}_{\mathrm{first\,SFR}} - t^{\ast}_{\mathrm{merge}}}
      dt_{\mathrm{delay}} \\
      \times \mathcal{N}_{\mathrm{form}}(Z, t_{\mathrm{delay}}, \zeta)
      \,\mathcal{S}(Z, z_{\mathrm{birth}})\,.
\end{split}
\label{eq:mrd}
\end{equation}

The inner integral is over the delay time, $t_{\rm delay}$, the time it takes from birth of the binary system to compact object merger. The upper limit is the difference between the time of the first star formation ($t^{\ast}_{\mathrm{first\,SFR}})$, based on the specific SFRD assumptions, and the merger time ($t^{\ast}_{\mathrm{merge}}$). 
We follow the notation from \citet{Hendriks:2023yrw} where a $*$ superscript indicates lookback time, while times without a $*$ superscript correspond to a duration. The outer integral over metallicity is evaluated over the metallicity range spanned by our simulations, see Section~\ref{subsec:compas} and Section~\ref{subsec:sevn} for each range, outside of this star-formation is taken to be zero by SSPC. The function $\mathcal{N}_{\mathrm{form}}(Z, t_{\mathrm{delay}}, \zeta)$ is the BBH formation efficiency, i.e. the number of BBH systems, $N_{\rm BBH}$, formed per solar mass:

\begin{equation}
    \mathcal{N}_{\mathrm{form}}(Z, t_{\mathrm{delay}}, \xi) = \frac{N_{\rm BBH} (Z, t_{\rm delay}, \xi)}{M_{\rm total, formed}} \,
    \label{eq:Nform}
\end{equation}

and is dependent on the metallicity, $Z$, delay time, $t_{\rm delay}$, and system properties, $\zeta$, e.g. component masses and orbital parameters. The `true' amount of star-forming mass, $M_{\rm total, formed}$, corrects for single stars and binaries formed outside our sampled mass range by analytically integrating the initial mass function (IMF) \citep{Kroupa2001} over the full mass range, following the renormalization scheme in \citet{Bavera:2020inc} (see their Appendix A), and assuming a binary fraction of unity throughout. 

The star-formation rate, $\mathcal{S}(Z, z_{\mathrm{birth}})$ is defined as a function of $Z$ and birth redshift, $z_{\rm birth}$. It is the combination between the star-formation rate density and the redshift-dependent metallicity distribution:

\begin{equation}
    \mathcal{S}(Z, z) = \mathrm{SFRD}(z) \times \frac{dP}{dZ}(Z,z) \,,
\end{equation}
where we use the functional form from \citet{Madau:2014bja} for our $\mathrm{SFRD}(z)$:

\begin{equation}
    \mathrm{SFRD}(z) = a\frac{\,(1+z)^{b}}{1 + [(1+z)/c]^{d}} \,,
\end{equation}

and the following  parameters $(a, b, c, d) = (0.02, 1.48, 4.45, 5.90)$ from \citet{vanSon:2021zpk}. The metallicity distribution, $\frac{dP}{dZ}(Z,z)$, follows \citet{vanSon:2022ylf}:

\begin{equation}
    \frac{dP}{dZ}(Z,z) = \frac{2}{Z}\, 
    \phi\!\left(\frac{\ln Z - \xi(z)}{\omega(z)}\right)
    \Phi\!\left(\alpha\,\frac{\ln Z - \xi(z)}{\omega(z)}\right),
\end{equation}

with $\phi$ the standard log-normal distribution, and $\Phi$ the cumulative distribution function of the standard log-normal distribution. The width is taken to be $\omega(z) = \omega_0 \cdot 10^{\omega_z \cdot z}$, with $\omega_0 = 1.125$ and $\omega_z = 0.05$. The function $\xi(z)$ is given as:

\begin{equation}
    \xi(z) = \ln \left( \frac{\mu_0 \cdot 10^{\mu_z \cdot z}}{2 \Phi(\beta \,  \omega (z))} \right) - \frac{\omega (z)^2}{2} \,,
\end{equation}

with $\mu_0=0.025$ and $\mu_z = -0.05$ and the parameter $\beta = \alpha/\sqrt{1+\alpha^2}$ with $\alpha = -1.77$.

This calculation is carried out in \texttt{SSPC} for both \texttt{COMPAS} and \texttt{SEVN}, providing a standardized calculation of the merger rate density and its derivatives.

\subsection{Selecting Mass Ratio Reversal Candidates}
Throughout, we will use the criteria for MRR that the mass of the black hole descending from the initially less massive star (B) needs to be larger than the initially more massive star's black hole (A), i.e. M$_{\rm B}$ > M$_{\rm A}$. While this includes systems which barely pass the threshold, we show in Appendix~\ref{app:mrr-threshold} that our results remain largely the same when imposing a conservative threshold of  M$_{\rm B}$ > 1.05 $\times$  M$_{\rm A}$. As we will show, MRR predictions differ substantially across the two population synthesis results, with discrepancies going beyond those found by varying this threshold.

\subsubsection{Observations}

We compare our population synthesis results with population models based on gravitational-wave observations from the LVK collaboration's GWTC-5 \citep{gwtc5-population}. In particular, we compare to the inferred primary-mass, secondary-mass, and mass-ratio distributions from the \textit{PixelPop} model and provide an additional comparison in the joint \mone--\mtwo space.

\section{Results}
\label{sec:results}

\begin{figure*}[t]
    \centering
    \includegraphics[width=0.7\linewidth]{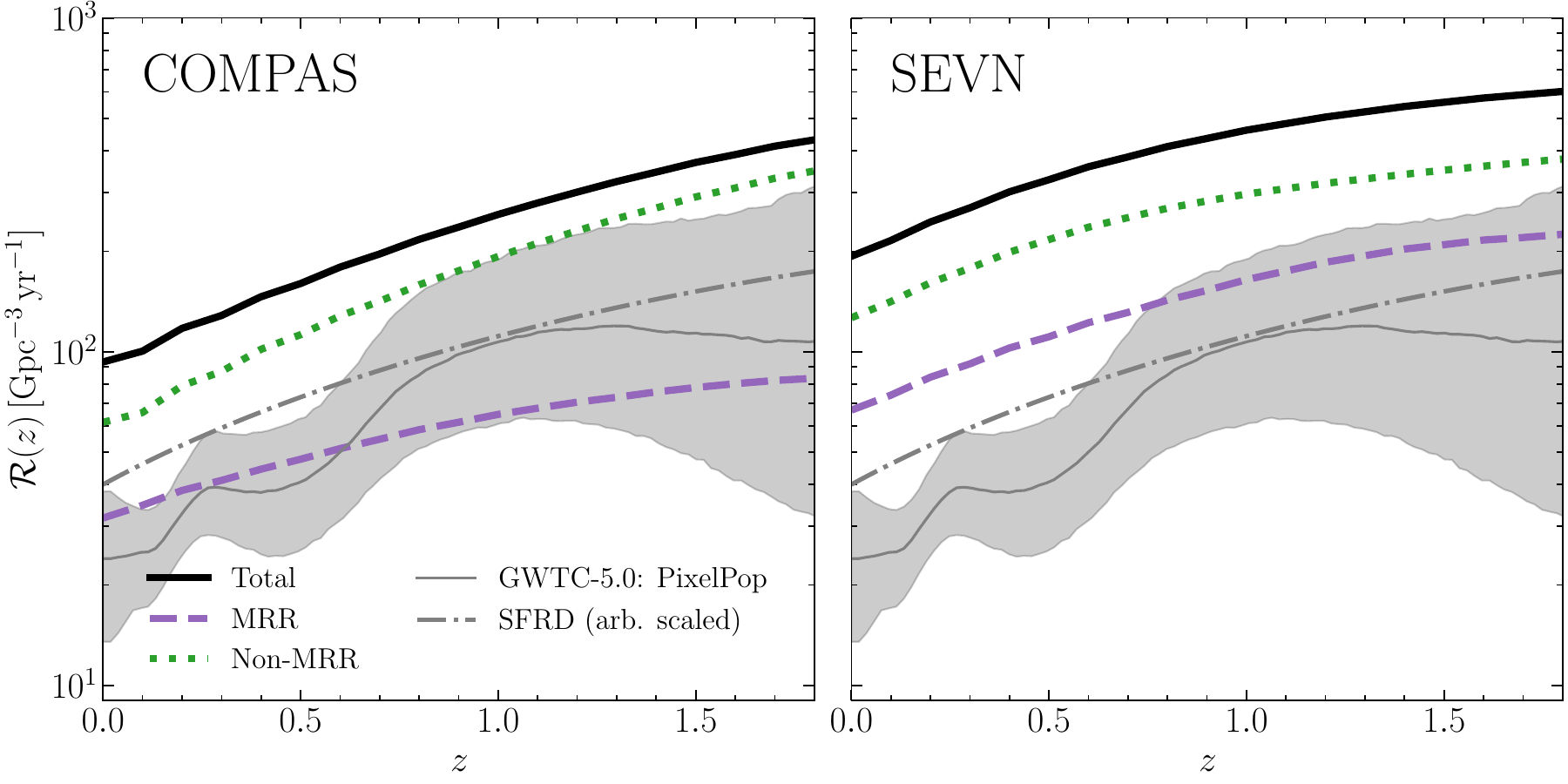}
    \caption{Redshift evolution of the BBH merger rate density $\mathcal{R}(z)$ expected by the fiducial \texttt{COMPAS} (left) and \texttt{SEVN} (right) models. 
    The total merger rate (black) is decomposed into contributions from MRR (purple) and non-MRR (green) systems, and compared to the LVK O4b \textit{PixelPop} population model (gray band; \citealt{gwtc5-population}). 
    In both models, MRR systems contribute approximately one-third of the total BBH merger rate across cosmic time, with only weak redshift evolution. 
    While the overall normalization differs from LVK constraints, both models broadly track the rise and fall of the cosmic star formation history (gray dashed-dotted line).}
    \label{fig:drdz_mrr}
\end{figure*}

\begin{figure*}[t!]
    \centering
    \includegraphics[width=\linewidth]{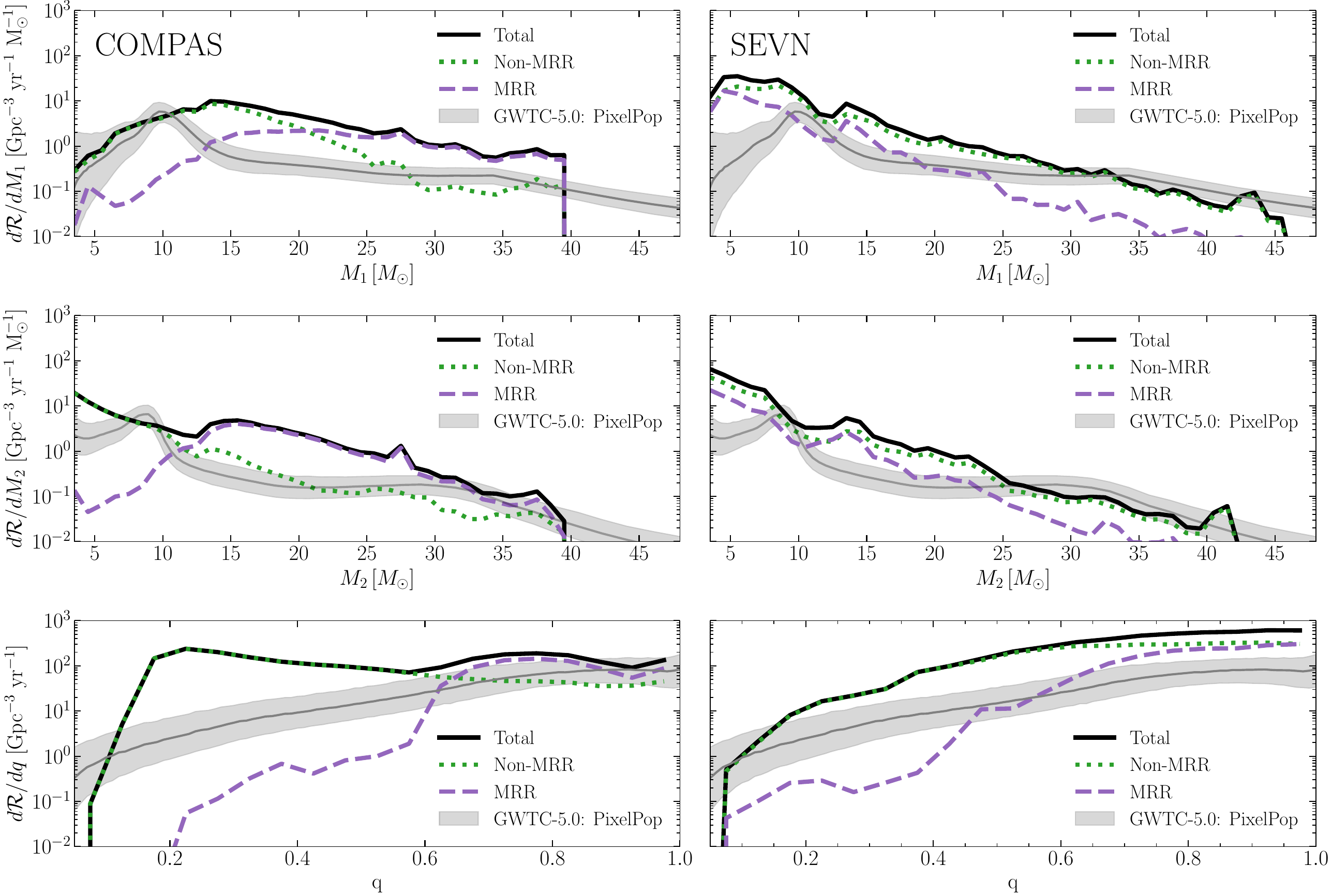}
    \caption{
    Intrinsic BBH merger rate density as a function of primary mass $M_1$ (top), secondary mass $M_2$ (middle), and mass ratio $q$ (bottom) at $z \approx 0.2$, for \texttt{COMPAS} (left) and \texttt{SEVN} (right). 
    The total population (black) is decomposed into MRR (purple) and non-MRR (green) contributions, and compared to LVK-inferred constraints from PixelPop, GWTC-5.0 (gray shaded regions; \citealt{gwtc5-population}). 
    \texttt{COMPAS} predicts that MRR systems dominate the high-mass regime ($M_1 \gtrsim 20 \Msun$, $M_2 \gtrsim 12 \Msun$), whereas \texttt{SEVN} predicts a more diffuse contribution that remains subdominant across the mass range. 
    In contrast, both models show that MRR preferentially populates the high-mass-ratio regime ($q \gtrsim 0.6$).}
    \label{fig:mass-distributions-MRR-vs-nonMRR}
\end{figure*}

\subsection{Imprint of Mass Ratio Reversal on BBH Merger Rates and Mass Distributions}
\label{subsec-results-imprint-MRR-on-BBHs}

We investigate how MRR imprints on both the overall BBH merger rate and the distribution of BBH masses. 
The evolutionary channels underlying MRR are discussed in Section~\ref{subsec:MRR-channels}.
Across both population synthesis frameworks, MRR constitutes a substantial fraction of merging systems, contributing to $\sim 1/3$ of the total merger rate. 
However, its impact on observable BBH properties differs qualitatively between models: \texttt{COMPAS} predicts that MRR systems dominate the high-mass and high-mass-ratio regimes ($q \gtrsim 0.6$), whereas \texttt{SEVN} predicts a more diffuse contribution that remains subdominant across the whole mass range. 
In this section, we first examine the redshift evolution of the merger rate and then quantify how MRR shapes the primary-mass, secondary-mass, and mass-ratio distributions.

\begin{figure*}[t!]
    \includegraphics[width=\linewidth]{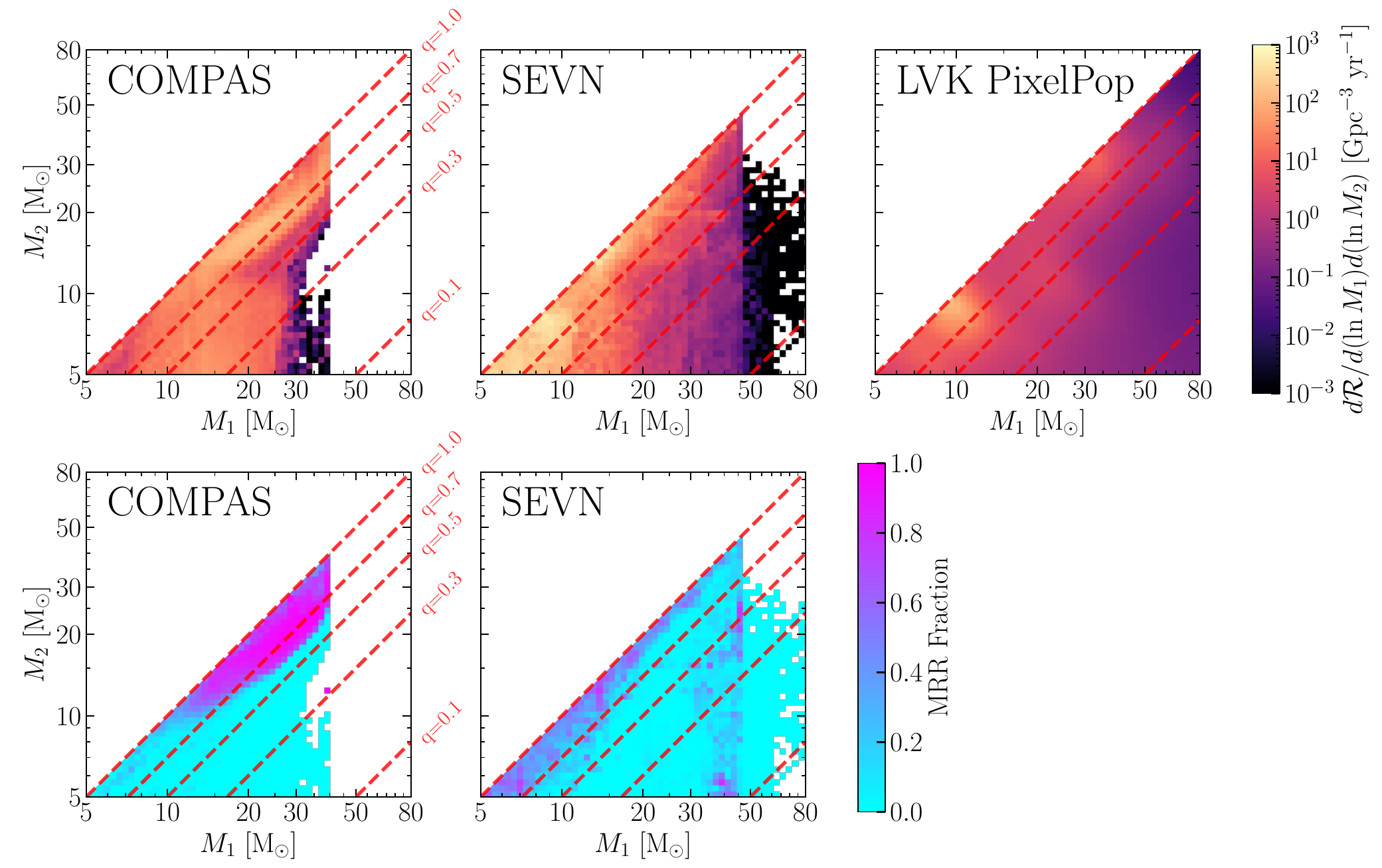}
    \caption{
        Two-dimensional BBH merger rate density ($z \approx 0$) in the $(M_1, M_2)$ plane (top row) and corresponding MRR fraction (bottom row) for \texttt{COMPAS} (left) and \texttt{SEVN} (middle), compared to LVK's PixelPop model (right). 
        Contours indicate lines of constant mass ratio $q$. 
        \texttt{COMPAS} exhibits a concentrated region at high masses and $q \gtrsim 0.6$ where MRR systems dominate, while \texttt{SEVN} shows no comparable region of dominance and instead displays a more diffuse MRR contribution across parameter space. 
        This contrast highlights that MRR can either restructure the high-mass BBH population (\texttt{COMPAS}) or remain a subdominant component (\texttt{SEVN}), despite similar overall MRR fractions.}
    \label{fig:m1m2_2D}
\end{figure*}

\subsubsection{MRR contribution to the merger rate}

Figure~\ref{fig:drdz_mrr} shows the expected total BBH merger rate density $\mathcal{R}_{\mathrm{BBH}}(z)$ for the fiducial \texttt{COMPAS} and \texttt{SEVN} models, decomposed into MRR and non-MRR contributions and compared to the LVK O4b \textit{PixelPop} population model \citep{gwtc5-population}. 
In both frameworks, the total merger rate rises toward $z \sim 1$--2, broadly tracking the cosmic star formation density (SFRD), while the contribution from MRR systems shows contrasting redshift evolution between the two codes: in \texttt{SEVN}, the MRR fraction remains approximately constant (rising mildly from $\sim35\%$ at $z=0$ to $\sim37\%$ by $z\approx1.8$), whereas in \texttt{COMPAS} it declines noticeably with redshift, from $\sim34\%$ at $z=0$ to $\sim21\%$ by $z\approx1.8$. Near $z\approx0.2$, where we focus our subsequent mass-distribution analysis, MRR systems contribute roughly one-third of the total merger rate in both models. The persistence of a substantial MRR fraction across this redshift range in both frameworks, despite their differing trends, indicates that the physical processes driving mass ratio reversal remain efficient across a broad range of metallicities and cosmic epochs, rather than being confined to a narrow formation environment; the differing slopes instead point to code-dependent sensitivities in how MRR responds to the redshift-evolving metallicity distribution, which we return to in Section~\ref{app:drdz-mrr-evolution}.

At $z\approx0.2$, \texttt{COMPAS} expects a local merger rate of $R_{\mathrm{BBH}} = 117~\mathrm{Gpc}^{-3}\,\mathrm{yr}^{-1}$, with $34\%$ arising from MRR systems, while \texttt{SEVN} predicts $R_{\mathrm{BBH}} = 244~\mathrm{Gpc}^{-3}\,\mathrm{yr}^{-1}$, with a similar MRR fraction of $\sim 33\%$. 
These values exceed the LVK-inferred range of $R_{\mathrm{BBH}} = 27.5$--$49.4~\mathrm{Gpc}^{-3}\,\mathrm{yr}^{-1}$ for the local BBH merger rate, built from the union of their \textit{FullPop} and \textit{PixelPop} models \cite{gwtc5-population}. 
This reflects known sensitivities of population synthesis predictions to assumptions such as the cosmic star formation history, metallicity evolution, and binary fraction that can impact the \textit{magnitude} of the BBH merger rate by orders of magnitude \citep[e.g.][and references therein]{Neijssel2019, Broekgaarden:2022imbII, vanSon:2022ylf, Santoliquido:2020bry, Chruslinska:2022ovf, MandelBroekgaarden:2021, vanSon2023, Romagnolo:2023, Sgalletta:2024,Romagnolo:2025,Sgalletta_2026}. 
Our focus in this work is therefore primarily on the \emph{relative} contribution of MRR systems and their imprint on the BBH mass distributions, rather than on the absolute normalization of the merger rate.

\subsubsection{MRR contribution to BBH masses}
\label{subsec:MRR-contribution-to-BBH-masses}

Figure~\ref{fig:mass-distributions-MRR-vs-nonMRR} shows the intrinsic BBH merger rate density as a function of component masses ($M_1, M_2$) and mass ratio $q = M_2/M_1$ at $z \approx 0.2$, decomposed into MRR and non-MRR contributions for the fiducial \texttt{COMPAS} and \texttt{SEVN} models. 
The primary and secondary masses are defined following the LVK convention, with $M_1 \geq M_2$. 
Figure~\ref{fig:m1m2_2D} shows the corresponding two-dimensional distribution of the MRR fraction in the $(M_1, M_2)$ plane. 
The two population synthesis frameworks predict qualitatively different roles for MRR in shaping the BBH mass distributions.

In the fiducial \texttt{COMPAS} model, MRR systems dominate the high-mass regime and reshape both component-mass distributions. 
The primary-mass distribution exhibits a peak near $15 \Msun$ dominated by non-MRR systems, while a secondary enhancement at $25$--$35 \Msun$ is driven by MRR systems, which dominate above $M_1 \gtrsim 20 \Msun$. 
A similar trend is present in the secondary-mass distribution, where MRR systems dominate for $M_2 \gtrsim 12 \Msun$. 
This behavior is even more apparent in the two-dimensional $(M_1, M_2)$ plane (Figure~\ref{fig:m1m2_2D}), where the MRR contribution is concentrated at high masses and high mass ratios ($q \gtrsim 0.6$).

In contrast, the fiducial \texttt{SEVN} model predicts that MRR systems remain subdominant across the full mass range and do not strongly shape the component-mass distributions. 
Instead, the MRR contribution peaks at lower masses ($M \lesssim 20 \Msun$) and is more diffusely distributed. 
In the $(M_1, M_2)$ plane, the MRR fraction is largest in the low-mass, high-$q$ region, without a clear regime where MRR dominates the overall population, aside from a preference for $q \gtrsim 0.7$. 
Combined, this demonstrates that, in \texttt{SEVN}, MRR contributes to but does not restructure the observable BBH mass distribution. 
These results demonstrate that MRR either restructures the high-mass BBH population (\texttt{COMPAS}) or remains a secondary contribution (\texttt{SEVN}), highlighting a fundamental model-dependent uncertainty in linking massive BBHs to their formation channels.

Despite these differences, both models predict that MRR systems preferentially populate the high-mass-ratio regime. 
In \texttt{COMPAS}, MRR systems dominate for $q \gtrsim 0.6$, while in \texttt{SEVN} they become most prominent at even higher mass ratios ($q \gtrsim 0.7$), though never dominating.
This reflects the tendency for MRR to arise in binaries that evolve toward near-equal-mass configurations through stable mass transfer (SMT) (see Section~\ref{subsec:MRR-in-sims} for more details). 
As a result, the high-$q$ regime consistently shows an enhanced MRR contribution across both population synthesis frameworks, even as its impact on the mass distributions differs substantially between them. At $q > 0.7$, MRR systems contribute $\sim 72\%$ of the merger rate in \texttt{COMPAS} and $\sim 44\%$ in \texttt{SEVN}; for mass ratios approaching unity ($q>0.9$), this gap narrows, with \texttt{COMPAS} at $\sim 64\%$ and \texttt{SEVN} approaching parity at $\sim 48\%$. This convergence suggests the high-$q$ regime is a promising region to search for MRR.

Taken together, these results show that while the overall fraction of MRR systems is similar between models, their observational imprint depends sensitively on how MRR systems populate the BBH mass distribution. 
This is reflected in the comparison to LVK-inferred mass distributions: \texttt{COMPAS} overpredicts intermediate-mass systems, while \texttt{SEVN} exhibits an excess at low masses. 
These discrepancies highlight that the imprint of MRR on the BBH population is intertwined with broader modeling uncertainties in stellar and binary evolution.

\begin{figure*}[t!]
    \centering
    \includegraphics[width=0.9\linewidth]{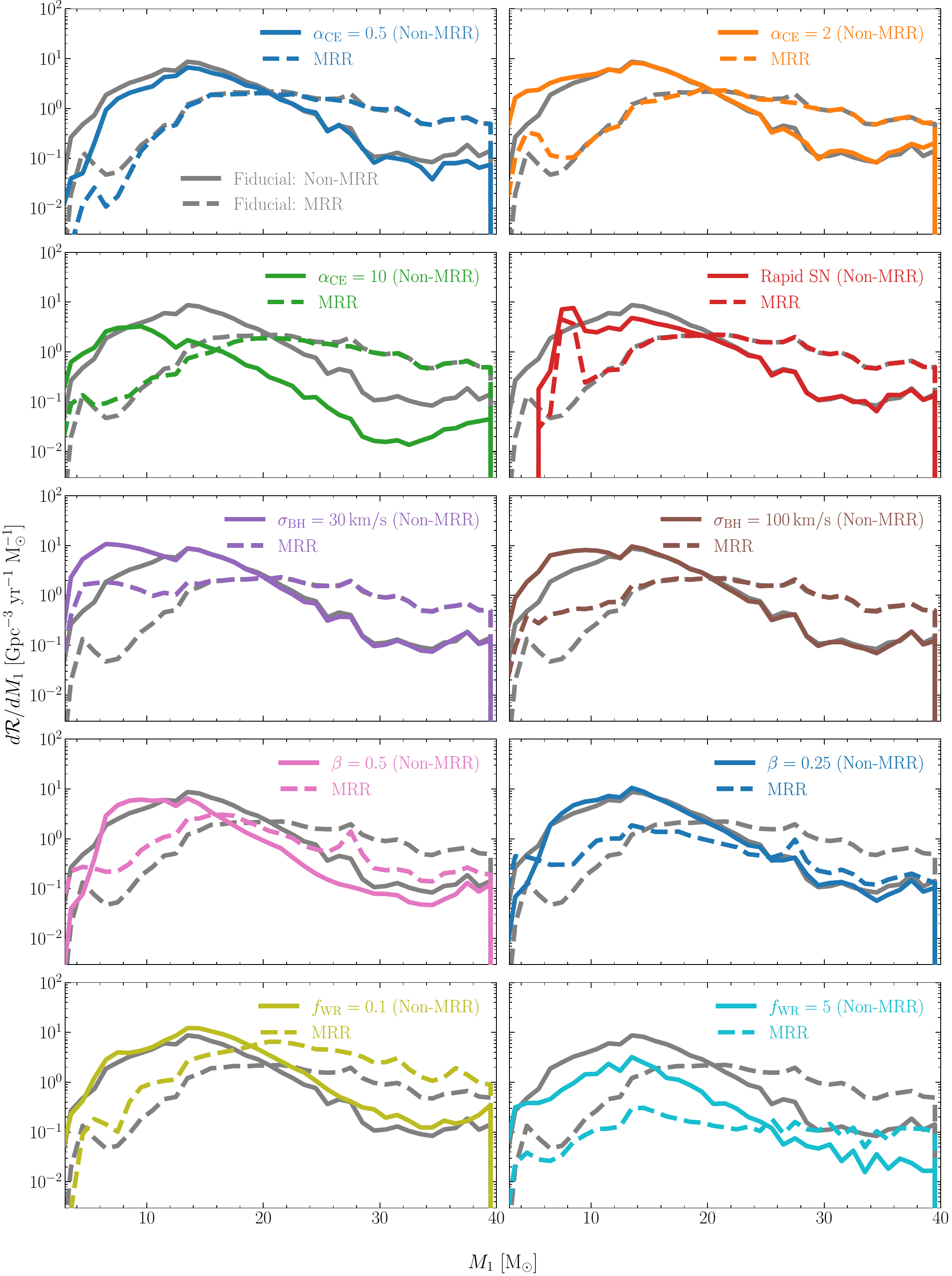}
\caption{
Sensitivity of the primary black hole mass distribution, $M_1$, to key uncertainties in massive binary evolution within \texttt{COMPAS}. 
Each panel shows a single-parameter variation relative to the fiducial model, with the $M_1$ distribution decomposed into MRR (dashed) and non-MRR (solid) contributions. 
Variations in the common-envelope efficiency $\alpha_{\rm CE}$, supernova engine prescription, and black-hole natal kick dispersion $\sigma_{\rm BH}$ primarily affect the low-mass regime ($M_1 \lesssim 15$--$20 \Msun$) by altering binary survival and compact-object formation pathways. 
In contrast, changes in the mass-transfer efficiency $\beta$ and Wolf--Rayet wind strength $f_{\rm WR}$ directly impact core growth and modify the MRR contribution across the full mass range. 
Despite these differences, all models consistently show that MRR systems preferentially populate the high-mass regime, demonstrating that this is a robust feature of the simulations. }
\label{fig:compas-variations}
\end{figure*}

\subsubsection{Impact of uncertainties in massive binary evolution in COMPAS}
\label{sec:results-impact-uncertainties-compas}

To assess the robustness of the trends identified above, we examine how uncertainties in key binary-evolution prescriptions affect the simulated BBH primary-mass distribution. 
We focus on single-parameter variations within \texttt{COMPAS}, where extensive public model grids enable a controlled exploration of physical assumptions (see Appendix~\ref{app:chirp-mass} for a complementary analysis of the chirp-mass distribution). 
The explored variations span physically motivated ranges relevant for BBH formation \citep{Broekgaarden:2022imbII}, including the common-envelope efficiency $\alpha_{\rm CE}$, supernova engine prescription, black-hole natal kick dispersion $\sigma_{\rm BH}$, mass-transfer efficiency $\beta$, and Wolf--Rayet wind strength $f_{\rm WR}$.

Figure~\ref{fig:compas-variations} shows that these parameters affect the primary-mass distribution in qualitatively different ways, reflecting the distinct physical processes they regulate. 
Variations in $\alpha_{\rm CE}$, $\sigma_{\rm BH}$, and the supernova engine primarily impact the low-mass end of the distribution ($M_1 \lesssim 15$--$20 \Msun$), below the regime in which MRR systems dominate. 
These parameters regulate binary survival and compact-object formation pathways, altering the normalization and shape of the low-mass spectrum while leaving the high-mass (MRR-dominated) regime largely unchanged. 
An exception occurs for extreme common-envelope efficiencies (e.g.\ $\alpha_{\rm CE}=10$), where the non-MRR contribution is significantly reduced while the MRR contribution remains comparable. Higher values of $\alpha_{\rm CE}$ lead to larger orbital separations, thus systems which already had wide separations end up not being able to merge in a Hubble time; considering MRR systems require close separations to initiate mass transfer, they may take longer to merge overall but are largely unaffected in this model.

In contrast, variations in the mass-transfer efficiency $\beta$ and Wolf--Rayet wind strength $f_{\rm WR}$ directly affect the growth and retention of the secondary’s mass and helium core prior to collapse. 
As a result, these parameters modify the MRR contribution across the full mass range: lower mass-transfer efficiency or stronger winds suppress MRR, while more efficient mass transfer enhances it. 
This highlights the central role of envelope stripping and core growth in setting the prevalence of MRR.

Despite these quantitative differences, all explored variations consistently show that MRR systems preferentially populate the high-mass regime, though the magnitude can vary with parameters, e.g.  $\beta$ and $f_{\rm WR}$. 
This indicates that the emergence of MRR at high primary masses is not driven by any single parameter choice within the ranges explored here, but instead represents a robust feature of isolated binary evolution in \texttt{COMPAS} across this parameter space.

\subsection{What it takes to MASSquerade}
\label{subsec:MRR-in-sims}
\begin{figure*}[t!]
    \centering
    \includegraphics[width=\linewidth]{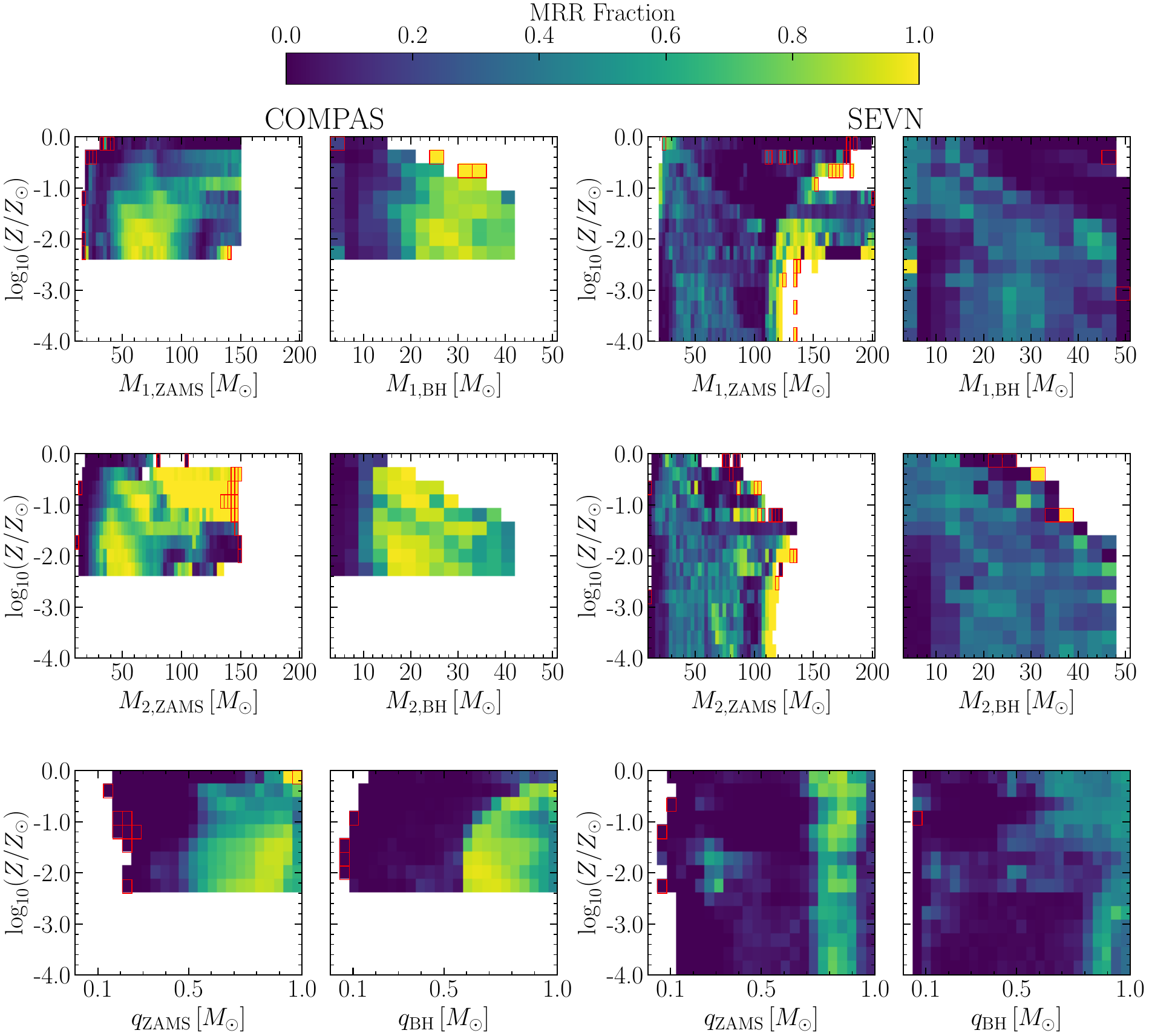}
    \caption{Fraction of systems undergoing mass ratio reversal (MRR) for \texttt{COMPAS} (left two columns) and \texttt{SEVN} (right two columns), weighted by their merger rate contribution at $z\approx 0.2$. White bins contain no systems contributing to the merger rate at this redshift. The red-framed bins indicate bins with fewer than 10 systems contributing to the merger rate. For each population synthesis code, the leftmost column shows the MRR fraction in the plane of ZAMS quantities and metallicity, while the right panels show the corresponding BH masses in the "LVK" defined view, i.e. \mone defined as the most massive BH and \mtwo defined as the least massive. Note the metallicity range of each simulation matches to its simulated grid (i.e. \texttt{SEVN} extends to lower metallicities). See Figure~\ref{fig:ZAMS-BBH-properties-binary-M1M2} for the mapping in the "binary" defined view, i.e. \mone is the most massive star at ZAMS. 
    }
    \label{fig:mrr_masses}
\end{figure*}

The results of Section~\ref{subsec-results-imprint-MRR-on-BBHs} show that MRR can either dominate the high-mass BBH population (\texttt{COMPAS}) or remain a subdominant contribution (\texttt{SEVN}), despite contributing a similar overall fraction of systems in both models. This raises a key question: which binary systems undergo mass ratio reversal, and what physical processes determine where they appear in the BBH mass distribution?
To address this, we connect the observable imprint of MRR to its progenitor properties by examining both the initial (ZAMS) parameter space and the evolutionary pathways that lead to MRR systems. 
We first map where MRR systems originate in ZAMS mass--metallicity space and how the less massive star can come to form the more massive BH, an effect we refer to as \textit{massquerading}: the initially less massive (secondary) star ``masquerades'' as the primary in the observed BBH mass distribution. We then identify the dominant formation channels that produce these systems.

\subsubsection{Origin of MRR systems in initial (ZAMS) parameter space}
\label{subsec:mrr-origins}
Figure~\ref{fig:mrr_masses} shows the MRR fraction as a function of component masses and metallicity, for both zero-age main sequence (ZAMS) binaries and their resulting black hole masses. Here M$_{1,\rm BH}$ is defined as the more massive black hole at merger, consistent with LVK's convention. A complementary version of this figure is presented in Appendix~\ref{sec-appendix-ZAMS-BBH-properties-binary-M1M2} with the direct mapping between the ZAMS masses and their corresponding BH masses. The fraction is weighted by the intrinsic BBH merger rate to reflect the progenitor properties of the underlying BBH merger population at $z\approx0.2$. In this subsection, we focus on the intrinsic formation pathways of MRR systems within the simulations.

In the fiducial \texttt{COMPAS} model, MRR systems are concentrated primarily in binaries with $50 \Msun \lesssim M_{1,\rm ZAMS} \lesssim 90 \Msun$ and metallicities $\log_{10}(Z/Z_\odot)\simeq-2.5$ to $-1$, with a secondary peak around $M_{1,\rm ZAMS} \sim 130 \Msun$ at lower metallicities $\log_{10}(Z/Z_{\odot}) \lesssim -1.6$. 
These systems span a broad range of secondary masses, $M_{2,\rm ZAMS}\sim30$--$140 \Msun$, but preferentially originate from binaries with initial mass ratios $0.5 \lesssim q_{\rm ZAMS}\lesssim 0.9$, reflecting that MRR is favored in systems that begin with comparable stellar masses.


While the mass reversal itself is driven by SMT growing the secondary's core (Section~\ref{subsec:MRR-channels}), successful MRR systems must also merge within a Hubble time, which requires sufficient orbital tightening. {At higher metallicity, wind-widened orbits are less likely to shrink enough during the initial stable mass transfer phase alone; additional orbital shrinkage can instead occur during a subsequent common-envelope phase. This second, orbit-tightening phase is preferentially triggered in systems with more massive secondaries and higher mass ratios, since these more evolved secondaries are more likely to fill their Roche lobe and initiate this interaction, which is why higher-metallicity MRR systems in Figure~\ref{fig:mrr_masses} skew toward higher initial mass ratios.

The MRR binaries predominantly produce black holes in the mass range $20$--$40 \Msun$, consistent with the regime where MRR dominates the BBH mass distribution (Section~\ref{subsec:MRR-contribution-to-BBH-masses}). 
Although higher metallicity systems tend to favor more equal-mass binaries, their overall contribution to the BBH population remains small due to the suppressed BBH formation efficiency at $\log_{10}(Z/Z_\odot)\gtrsim-0.5$.


In contrast, \texttt{SEVN} predicts that MRR is most prominent in the high-mass population, which is concentrated at $M_{1,\mathrm{ZAMS}} \gtrsim100 \Msun$ across a broad metallicity range $-4 \lesssim \log_{10}(Z/Z_\odot)\lesssim-0.4$, corresponding to systems that approach or enter the pair-instability regime, where strong pulsational mass loss significantly alters the final remnant mass (see Section~\ref{subsec:MRR-channels}). At high masses, there are less overall systems (IMF suppression) and a higher MRR fraction is thus easier to achieve.
A second population of MRR systems is present at lower primary masses and preferentially higher initial mass ratios, similar to systems found in \texttt{COMPAS}. 

However, unlike in \texttt{COMPAS}, these two populations map differently to remnant masses. The high-mass systems experience substantial mass loss due to pulsational pair-instability and have lower overall contribution due to IMF suppression, while the lower-mass, high-$q$ systems produce BHs spanning a broad mass range in \texttt{SEVN}, in contrast to the comparatively narrow remnant-mass range these systems produce in \texttt{COMPAS}. This broader mapping likely reflects differences in how \texttt{SEVN} handles mass-transfer efficiency relative to \texttt{COMPAS}'s prescription. Regardless of its origin, the effect of this dispersion is that no single BH-mass range is dominated by MRR systems in \texttt{SEVN}, which explains why MRR does not restructure the BBH mass distribution in this model, despite contributing a non-negligible fraction of mergers.

The MRR fraction shown in this section identifies where MRR is most \textit{efficient}, i.e. where a large share of the systems in a given bin undergo MRR, rather than where the bulk of the merger-rate-weighted MRR population originates. For example, in \texttt{COMPAS} only $\sim5\%$ of the total rate comes from bins with MRR fraction $\geq0.7$, while $\sim43\%$ comes from bins with fraction $\geq0.3$; regions of very high MRR fraction can therefore correspond to metallicity--mass combinations that are rare in the overall population. The rate-weighted marginal distributions in Figure~\ref{fig:mass-distributions-MRR-vs-nonMRR} ($d\mathcal{R}/dM_1$, $d\mathcal{R}/dM_2$, $d\mathcal{R}/dq$) remain the correct diagnostic for quantifying how much MRR contributes to the observable merger rate.

Furthermore, we note that the \texttt{SEVN} simulation grid extends to higher $M_{\rm 1, ZAMS}$ and lower metallicity than our \texttt{COMPAS} simulations, restricting \texttt{SEVN} to the \texttt{COMPAS} range has a negligible effect on the overall merger rate density and mass distributions, but would exclude the high-mass, PPISN population described above. We therefore retain the full \texttt{SEVN} range throughout.

\subsubsection{Formation channels of MRR systems}
\label{subsec:MRR-channels}

The trends and structures in initial and remnant parameter space indicate that MRR arises from multiple evolutionary pathways and a range of initial conditions rather than a single formation channel. We identify three distinct channels through which the initially less massive star can ultimately form the more massive black hole: \textit{core growth via stable mass transfer}, \textit{PPISN-induced shrinking}, and an \textit{asymmetric-CCSN} channel; within the dominant core-growth channel, we further identify a PPISN-assisted variant discussed below. We include the temporal evolution of the total and core masses for the evolutionary channels in Figure~\ref{fig:mrr_channels}. These channels occupy distinct regions of initial parameter space, as shown in Figure~\ref{fig:mrr_mass_channel_MRD}.

\begin{figure*}
    \centering
    \includegraphics[width=0.8\linewidth]{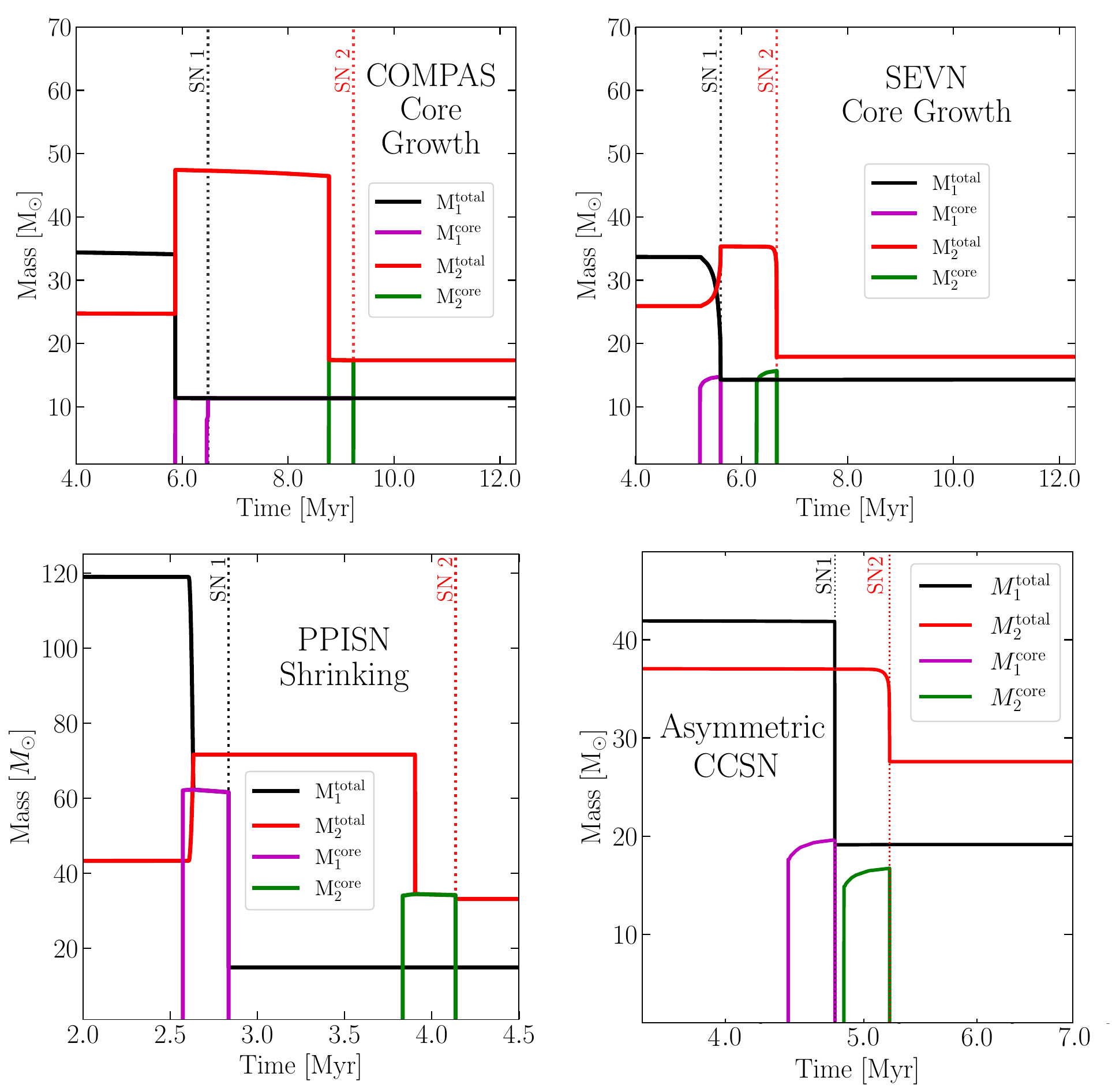}
    \caption{
Representative evolutionary pathways leading to MRR in binary systems. 
Shown are the total stellar masses of the initially more massive star (black) and initially less massive star (red), together with their corresponding core masses (magenta and green), as a function of time. 
Vertical dotted lines mark the supernova events of the primary (SN1) and secondary (SN2).
\textbf{Top left:} Core-growth channel (\texttt{COMPAS}). Stable mass transfer during post–main-sequence evolution increases the helium-core mass of the secondary, leading to a more massive black hole despite its initially lower ZAMS mass. The reversal here is driven by secular core growth rather than impulsive mass loss.
\textbf{Top right:} Core-growth channel (\texttt{SEVN}). The same as the first panel, with a similar system, but evolved in \texttt{SEVN}. We note that in \texttt{SEVN} the core growth is a gradual process with the primary losing less mass overall. This highlights the differences in stellar evolution mentioned in the text.
\textbf{Bottom left:} PPISN-shrinkage channel. Strong pulsational pair-instability episodes strip the primary near the PPISN–PISN boundary, reducing its final remnant mass below that of the secondary.
\textbf{Bottom right:} Asymmetric-CCSN channel. The primary becomes stripped while the secondary retains its Hydrogen envelope, leading to the primary star becoming the less massive black hole.  
 }
\label{fig:mrr_channels}
\end{figure*}

\paragraph{\textit{Core growth via stable mass transfer channel}}
The dominant pathway in both \texttt{COMPAS} and \texttt{SEVN} is a channel we identify as \textit{core growth via stable mass transfer}. 
In this channel, top-left panel in Figure~\ref{fig:mrr_channels}, the initially more massive star transfers mass after leaving the main sequence, while the secondary accretes (part of) the transferred material while still on the main sequence. 
This accretion phase allows the secondary to significantly grow its core mass, such that by the time both stars have undergone core collapse, the initially less massive star can form the more massive black hole. The mass-ratio reversal is therefore  primarily driven by the core growth during stable mass transfer. While similar initial parameters were chosen for the systems in Figure~\ref{fig:mrr_channels}, it is clear that the two population synthesis codes treat stellar evolution differently. In \texttt{SEVN} (top-right panel) the cores are grown gradually with the secondary growing to a similar size as in \texttt{COMPAS}, while the primary loses less overall mass leading to a higher BH mass ratio. Overall, when weighted by the intrinsic BBH merger rate, this pathway accounts for $\sim99\%$ of all MRR systems in both codes, reflecting its dominant contribution to the MRD at $z\approx0.2$ shown in Figure~\ref{fig:mrr_mass_channel_MRD}.

\paragraph{\textit{PPISN-shrinking channel}} 
A second channel opens in \texttt{SEVN}, in which the primary undergoes strong pulsations near the PPISN-PISN boundary, removing a substantial fraction of its mass prior to collapse and reducing its final remnant mass below that of the secondary (see the bottom-left panel of Figure~\ref{fig:mrr_channels}). 
The secondary, which avoids this strong pulsational mass-loss phase, experiences weaker PPISN stripping and therefore retains a larger final remnant mass, ultimately forming the more massive black hole.
We refer to this pathway as \textit{PPISN-induced shrinking}.  
Although this channel contributes less than $1\%$ of the MRD-weighted MRR systems, it dominates the MRR contribution at high $M_{1, \rm ZAMS}$ and low mass ratios. 
In contrast to the core growth channel, the mass ratio reversal here is driven primarily by asymmetric mass loss at late evolutionary stages. We note that it is likely that if the \texttt{COMPAS} runs included metallicities of $\log(Z/Z_\odot) < -2.5$, potential PPISN-shrinking systems would arise.

\paragraph{\textit{Asymmetric CCSN channel}}
The final pathway identified in \texttt{SEVN} is driven by asymmetric mass loss during the CCSN phase and contributes 1.5$\%$ of the MRD-weighted MRR systems. 
This channel is distinct from the core growth channel, where the reversal is encoded in the pre-SN core mass hierarchy, and from the PPISN shrinkage channel, where the reversal is driven by enhanced pulsational stripping of the primary's mass. 
In the asymmetric-CCSN channel, some primary stars experience envelope stripping prior to collapse, such that the primary is stripped while the secondary retains its envelope, leading to substantially different CCSN mass loss between the two stars (see bottom-right panel of Figure~\ref{fig:mrr_channels}). 
However, the channel also includes systems in which both stars are stripped, but the non-linear mapping between pre-supernova structure and remnant mass causes the secondary to lose relatively less mass  during collapse and ultimately form  the more massive remnant. 
Overall, this channel remains subdominant across the full initial parameter space. 

\paragraph{\textit{A core-growth variant: PPISN-assisted core growth}}
%
The small remaining fraction of systems in \texttt{COMPAS} follow an evolutionary pathway similar to the  core-growth channel described above, but with one important difference: the secondary's core mass  never exceeds that of the primary. 
Instead, stable mass transfer drives the system  toward nearly equal core masses, placing both stars in the pulsational-pair instability supernova (PPISN) regime. Because the primary still retains a slightly larger core,  it loses marginally stronger PPISN mass loss than the secondary, ultimately producing the less massive black hole. We refer to this pathway as \textit{PPISN-assisted core-growth}. Because the evolution is a mix between core-growth and PPISN-shrinking (with a much smaller mass-loss differential), we do not include a plot of its temporal evolution. This channel contributes only a small fraction ($\lesssim1\%$) of MRR systems and is found primarily at the lowest mass ratios. 

Throughout both codes, metallicity plays a secondary role with core-growth dominating entirely. The contributions of the other channels are minimal: PPISN-assisted core-growth truncates above $Z\gtrsim-1.3$ in \texttt{COMPAS}, a similar truncation occurs for PPISN-shrinking in \texttt{SEVN}, and asymmetric-CCSN follows the core-growth shape in \texttt{SEVN} but contributes minimally.

The dependence of MRR on orbital separation for both \texttt{COMPAS} and \texttt{SEVN} (including the evolutionary channels) is included in Appendix~\ref{app:orbital-sep}. 
 

Taken together, these results demonstrate that the presence and observational imprint of MRR systems are set primarily by the initial mass ratio and the efficiency of mass transfer during the core-growth channel, which accounts for $\sim99\%$ of MRR systems in both codes. A higher initial mass ratio makes it easier for the secondary to grow its core mass to meet or exceed that of the initially more massive primary, so MRR is favored in binaries that begin close to equal mass. The contrasting MRR populations predicted by \texttt{COMPAS} and \texttt{SEVN} therefore reflect differences in how each code treats mass-transfer efficiency rather than differences in the subdominant pair-instability or asymmetric-CCSN channels: as summarized in Section~\ref{subsec:defaults}, \texttt{COMPAS} adopts thermal-timescale-limited accretion set by each mass transfer episode, while \texttt{SEVN} adopts a fixed mass-transfer efficiency of $\beta=0.5$. These differing prescriptions govern how efficiently the secondary can accrete mass from the donor and grow its core. In \texttt{COMPAS}, the fiducial thermal-timescale prescription is comparable to a fixed $\beta=0.5$ at high masses, where MRR dominates, while $\beta=0.5$ additionally raises the MRR rate at lower masses, increasing the total MRR fraction (Figure~\ref{fig:compas-variations}, but also see Section~\ref{app:drdz-mrr-evolution} below). A fixed $\beta=0.25$ is comparatively inefficient and suppresses MRR across the mass range. Together, these differences in efficiency lead to a different mass and mass-ratio dependence of MRR in the two codes.

\begin{figure*}
    \centering
    \includegraphics[width=\linewidth]{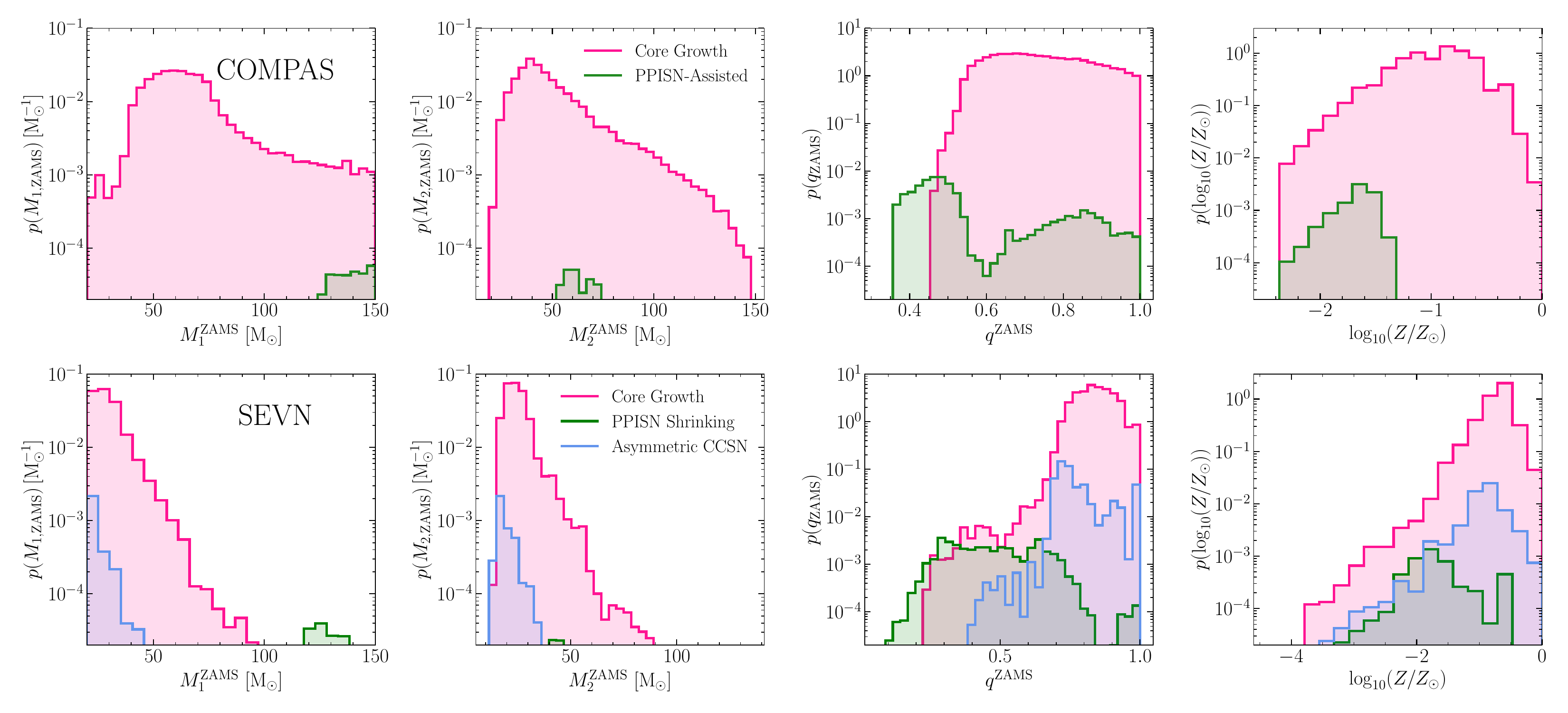}
    \caption{Initial parameter distributions of the distinct MRR channels in \texttt{COMPAS} (top-row) and \texttt{SEVN} (bottom-row). The distributions are normalized jointly across each codes channels, so that the area under the sum of all curves equals unity per panel. From left to right the columns are: ZAMS primary mass, ZAMS secondary mass, initial mass ratio, and metallicity. The channels for \texttt{COMPAS}  are core-growth (pink) and PPISN-assisted core-growth (green), while for \texttt{SEVN} they are the PPISN-shrink (green), core-growth (pink), and asymmetric-CCSN (blue). 
    The core-growth channel accounts for $99\%$ of MRR systems dominating throughout the initial parameter space for both codes. In \texttt{COMPAS} the PPISN-assisted core-growth dominates only at low-mass ratios. While in \texttt{SEVN}, a small dominant contribution for PPISN-shrink arises at high $M_{1, \rm ZAMS}$ and small mass ratio. The asymmetric-CCSN channel in \texttt{SEVN} remains subdominant, peaking at high component masses and mass ratio.
    Appendix~\ref{app:mrrchannels-formation-eff} shows how the relative  contributions of these channels change when weighted by the formation efficiency (Eq.~\ref{eq:Nform}) instead of the merger rate density. }
\label{fig:mrr_mass_channel_MRD}
\end{figure*}

\section{Discussion}
\label{sec:discussion}

\subsection{Massquerading in LVK: when a secondary masquerades as a primary}

A central implication of our results is that \ac{MRR} fundamentally complicates the interpretation of observed \ac{BBH} populations. 
In particular, MRR enables black holes formed from the initially less massive star (the secondary) to appear as the more massive component ($M_1$) in the observed BBH system, the massquerading effect introduced in Section~\ref{subsec:MRR-in-sims}.

Considering the massquerading phenomena is important to understand features in the BBH space and map back to formation pathways of their progenitor stars. 
From the perspective of binary stellar evolution, the initially more massive star (the donor in the first mass-transfer phase) and the initially less massive star (the accretor) follow systematically different evolutionary pathways in the typical formation channel of a BBH source \citep{vanSon2023}. 
The donor evolves first, loses mass through winds and mass transfer, and forms the first compact object, whereas the accretor can gain mass, angular momentum, and build a more massive helium core prior to collapse. 
As a result, the two stars leave distinct imprints on the resulting black hole population, a phenomenon well studied in binary evolution studies (e.g.\ \citealt{KippenhahnWeigert:1967, Wellstein:2001, deMink:2013, Renzo:2019,Renzo:2023,  Renzo:2021, Bavera:2021, Wagg:2024, Sen:2026, Zapartas:2021, Zapartas:2026}).

This distinction is reflected in our simulations through the different mappings from ZAMS mass to black hole mass for primaries and secondaries (Figure~\ref{fig:mrr_masses}; see also Appendix~\ref{sec-appendix-ZAMS-BBH-properties-binary-M1M2}). 
Secondaries that undergo efficient accretion can develop disproportionately large cores and form black holes that exceed the mass of the primary remnant, despite their initially lower mass. 

However, gravitational-wave observations do not retain information about the progenitor identities of the two compact objects. 
Instead, the LVK collaboration defines $M_1$ and $M_2$ purely by rank ordering ($M_1 \geq M_2$), independent of their evolutionary origin. 
This creates a fundamental mismatch between theoretically meaningful labels (donor vs.\ accretor) and observational labels (more massive vs.\ less massive black hole)  \citep[e.g.][]{Broekgaarden:2022, ZevinBavera:2022, mould22dec}.

Figure~\ref{fig:Massquerading-in-LVK} illustrates this mismatch directly. 
When black holes are classified according to their mass ordering at ZAMS (``binary'' definition), the mass distribution separates into contributions from remnants of initially more massive stars and those formed from initially less massive stars. 
In \texttt{COMPAS}, these two populations occupy partially distinct regions of parameter space, reflecting their different evolutionary histories, and re-labeling according to the observational convention (``LVK'' definition) substantially mixes them: black holes originating from secondaries are reassigned into the $M_1$ distribution whenever they exceed the mass of the primary remnant. 
In \texttt{SEVN}, by contrast, the binary- and LVK-defined $M_1$ and $M_2$ distributions occupy comparable regions of parameter space even before relabeling, so the observational convention produces a comparatively modest reshuffling relative to \texttt{COMPAS}. 
This again reflects the broader, less mass-concentrated MRR population predicted by \texttt{SEVN} (Section~\ref{subsec:MRR-contribution-to-BBH-masses}). 
See Figure~\ref{fig:compas-variations-massquerade} in Appendix~\ref{app:massquerading-variations} for the effects of single parameter variations in \texttt{COMPAS} on massquerading.


As a result, the observed primary-mass distribution $dR/dM_1$ is not a direct tracer of the evolution of initially more massive stars, but instead represents a superposition of two physically distinct populations. 
In particular, features produced by accretor evolution—such as enhanced core growth through stable mass transfer—can appear as excess structure in the observed $M_1$ distribution. 
This effect is clearly visible in Figure~\ref{fig:Massquerading-in-LVK}, where the ``LVK'' $M_1$ distribution shows contributions at intermediate masses that originate from the secondary population in the binary-defined view.

This massquerading effect provides a natural explanation for several features of the \texttt{COMPAS} and \texttt{SEVN} mass distributions discussed in Section~\ref{subsec:MRR-contribution-to-BBH-masses}. In particular, in \texttt{COMPAS}, the excess of systems in the $\sim10$--$30 \Msun$ range arises in part from MRR systems in which the accretor forms the more massive black hole and is therefore counted as $M_1$ observationally. 
Conversely, the donor remnants are shifted into the $M_2$ distribution, altering both component-mass spectra simultaneously.

More broadly, this implies that the commonly used interpretation of $M_1$ as tracing the evolution of the initially more massive star is not valid in the presence of MRR. 
Instead, $M_1$ must be interpreted as a mixed population whose composition depends on the efficiency of mass transfer, wind mass loss, and core growth. 
Neglecting this mixing can therefore bias inferences on key aspects of massive binary evolution, including the strength of stellar winds, the efficiency of envelope stripping, and the mapping from stellar core mass to compact remnant mass or understanding which of the BHs is spinning (e.g.\ \citealt{Stevenson:2017, MandelBroekgaarden:2021, Zevin2021}).

The strength of the massquerading effect is model-dependent. 
In \texttt{COMPAS}, where MRR systems dominate the high-mass regime, secondaries contribute substantially to the observed $M_1$ distribution, leading to a pronounced reshaping of the BBH mass spectrum. 
In contrast, in \texttt{SEVN}, where MRR systems are more diffusely distributed in remnant mass, massquerading is less localized in parameter space and has little discernible effect on the shape of the observed mass distributions.

\begin{figure*}[t]
    \centering
    \includegraphics[width=0.9\linewidth]{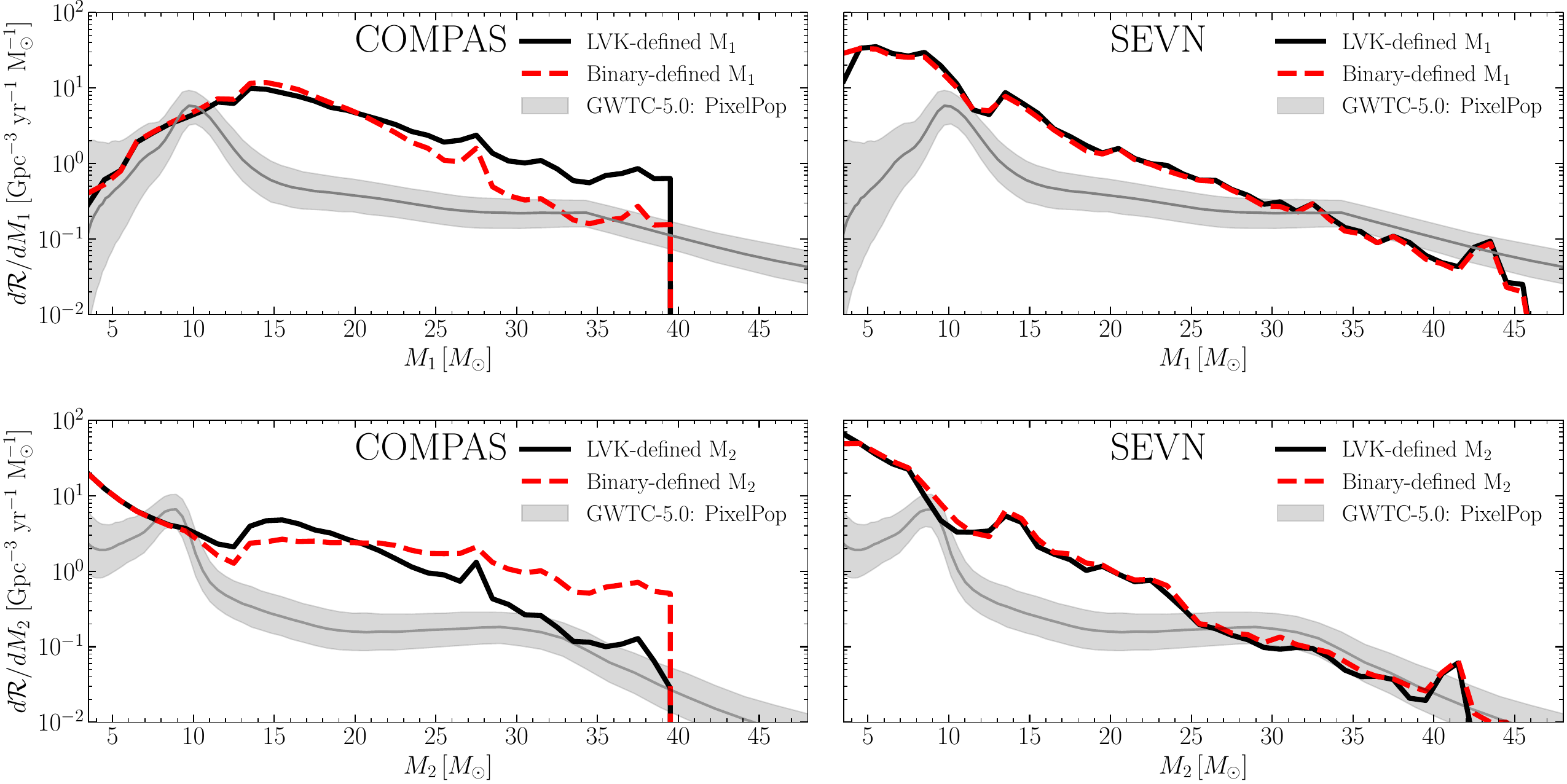}
    \caption{Massquerading in the component-mass distributions for \texttt{COMPAS} (left) and \texttt{SEVN} (right). The results from \texttt{COMPAS} show that depending on whether the primary black hole is classified based on being the larger black hole or stemming from the initially larger progenitor can have significant effects on the high-end regime in the mass distributions. Conversely, this definition has little effect on the results from \texttt{SEVN} because MRR never dominates a specific region in the mass distributions. The gray bands correspond to LVK-inferred constraints \citep{gwtc5-population}. }
    \label{fig:Massquerading-in-LVK}
\end{figure*}

Taken together, these results highlight that interpreting gravitational-wave observations requires careful consideration of the mapping between progenitor properties and observed quantities. 
In particular, forward-modeling approaches that retain information about progenitor identity, or inference frameworks that explicitly marginalize over formation channels, will be essential to disentangle the contributions of donor and accretor evolution in the observed BBH population.

\subsection{MRR fraction across cosmic time}
\label{app:drdz-mrr-evolution}

Figure~\ref{fig:drdz-mrr-evolution} shows the fraction of binary black hole mergers undergoing mass ratio reversal as a function of redshift. The fiducial \texttt{COMPAS} model is shown in black, while colored curves correspond to variations in key binary-evolution prescriptions, including the common-envelope efficiency ($\alpha_{\rm CE}$), black-hole natal kick dispersion ($\sigma_{\rm BH}$), the mass-transfer efficiency parameter ($\beta$), and the Wolf–Rayet wind mass-loss strength ($f_{\rm WR}$). Across all models, the MRR fraction decreases with increasing redshift. Early stars are metal poor, as such they lose less mass to stellar winds and ultimately end their lives as more massive black holes than their metal-rich counterparts. In order to undergo MRR, the less massive star needs to overcome the mass of the primary, which becomes increasingly difficult at early times. 

Notably, the $\alpha_{\rm CE}=10$ variation has the highest MRR fraction throughout cosmic time. This can be understood in conjunction with Figure~\ref{fig:compas-variations}, where the MRR contribution to the primary BBH mass distribution is largely unchanged, but the non-MRR contribution is greatly reduced leading to an overall increase in the total MRR fraction. For other values of $\alpha_{\rm CE}$ the MRR fraction doesn't deviate largely from the fiducial model.

The overall MRR fraction is particularly sensitive to the Wolf–Rayet wind mass-loss strength. In fact, the lowest MRR fraction across cosmic time is seen for the model with $f_{\rm WR}=5$. This is also clear from Figure~\ref{fig:compas-variations}, where the MRR contribution is largely reduced compared to the non-MRR contribution. The fiducial model, with $f_{\rm WR}=1$, doubles the MRR fraction at early times, and further lowering this parameter to $f_{\rm WR}=0.1$ increases the MRR fraction another $\sim70\%$ at early times. At early times, $z \gtrsim1.0$, these discrepancies lower to less than a factor of 2 following the metallicity dependence of WR formation and mass loss rates. 

The remaining model variations have modest departures from the fiducial model in terms of MRR fraction. The rapid supernova engine increases the overall MRR fraction $\sim30\%$ over the redshift range shown. The BH natal kick dispersion $\sigma_{\rm BH}$ affects the MRR fraction at the sub $\sim 15\%$ level, increasing the non-MRR and MRR fraction in tandem as can be seen in Figure~\ref{fig:compas-variations}. Finally, the shift to a constant mass-transfer efficiency of $\beta=0.5$ slightly increases the MRR fraction, while lowering it to $\beta=0.25$ suppresses the MRR fraction substantially. Overall, the MRR fraction's evolution consistently increases as time evolves and is mainly affected by winds, envelope stripping, and mass transfer.

In contrast to \texttt{COMPAS}, the fiducial \texttt{SEVN} model (black dashed line in Figure~\ref{fig:drdz-mrr-evolution}) shows a comparatively flat, mildly \textit{increasing} MRR fraction with redshift, rising from $\sim35\%$ at $z=0$ to $\sim37\%$ by $z\approx1.8$, in contrast to the fiducial \texttt{COMPAS} model, which declines from $\sim34\%$ to $\sim21\%$ over the same range. Decomposing the \texttt{SEVN} MRR population into its constituent channels (Section~\ref{subsec:MRR-channels}) shows that core growth remains the dominant channel at every redshift, accounting for $\gtrsim98\%$ of MRR systems from $z=0$ to $z=1.8$; the PPISN-shrink and asymmetric-CCSN channels each contribute at the sub-percent to few-percent level throughout and do not drive the small rise in the total MRR fraction. We verified that this contrast between codes is not an artifact of \texttt{SEVN}'s broader simulated mass and metallicity range: restricting the \texttt{SEVN} population to the same $M_{1,\rm ZAMS}$ and metallicity range simulated in \texttt{COMPAS} reproduces the unrestricted \texttt{SEVN} MRR fraction to within $\lesssim0.1 \%$ at every redshift tested. The differing redshift dependence between the two codes therefore reflects a genuine difference in how each code's core-growth channel responds to the redshift-dependent shift in typical progenitor metallicity, consistent with the differing mass-transfer prescriptions discussed in Section~\ref{subsec:defaults}, rather than a difference in which regions of initial parameter space each code samples.

\begin{figure}[h]
    \centering
    \includegraphics[width=\linewidth]{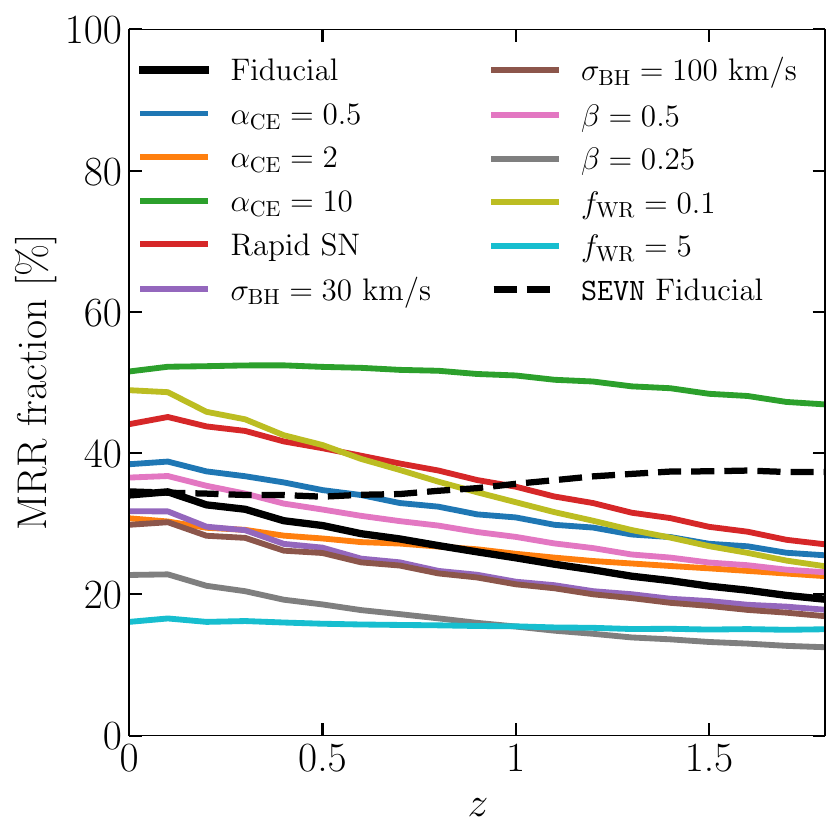}
    \caption{The mass ratio reversal contribution to the merger rate density across cosmic time. The black line is the fiducial \texttt{COMPAS} model, while the colored lines correspond to single-parameter variations in \texttt{COMPAS}; the black dashed line shows the fiducial \texttt{SEVN} model for comparison. Throughout cosmic evolution, all \texttt{COMPAS} variations show a decline in MRR fraction, hinting at a dependence on metallicity and/or delay time, while \texttt{SEVN} shows a mild increase instead, indicating that this trend is not universal across population synthesis codes.}
    \label{fig:drdz-mrr-evolution}
\end{figure}

\subsection{Comparison to earlier work}
Our results on MRR in BBH systems are broadly consistent with previous studies detailed here. Much of the existing literature has focused on the connection between MRR and BH spins, whereas here we emphasize its imprint on the mass distribution of BBHs. 

Using the same set of \texttt{COMPAS} simulations adopted in this work, \citet{Broekgaarden:2022} found that MRR occurs in $\gtrsim 70\%$ of BBHs observable by LVK and between 11--82$\%$ of the astrophysical population, depending on model assumptions. They showed that MRR occurs for masses $\gtrsim 20 \Msun$ and mass ratios of $\sim 0.7$ in agreement with our findings. They further demonstrated that the second-born BH can acquire non-negligible spin in up to $25\%$ of BBH systems and that the MRR systems can account for the anti-correlation in $\chi_{\rm eff}$--$q$ reinforced by later studies, e.g. \citet{Banerjee:2024}.

Similar conclusions were reached by \citet{ZevinBavera:2022}, who emphasized the role of MRR in producing asymmetric-mass BBH systems with a spinning primary BH. They found that the accretion efficiency during mass transfer is the dominant factor determining whether a system undergoes MRR. This is in direct agreement with our results (Section~\ref{sec:results-impact-uncertainties-compas}), where we find that higher mass transfer efficiency leads to a significantly larger fraction of MRR systems.

Consistent with this picture, \citet{Olejak:2021iux,Olejak:2024} showed that SMT naturally leads to MRR in BBH systems, typically producing mass ratios in the range $q \sim 0.4$--$0.7$. These systems can also give rise to primary black holes with non-negligible spins, driven by accretion and tidal spin-up during binary evolution, further reinforcing the connection between MRR and observable spin signatures.


Recently, on the observational side, \citet{Sen:2026} use samples of Algol-type binaries, i.e. systems having undergone MRR, to constrain mass transfer efficiency and angular momentum loss using an analytic framework. We note, however, that Algol binaries are generally understood to interact at comparatively tight orbital separations compared to the wider progenitor systems relevant for BBH formation in this work. Since mass-transfer efficiency is expected to decrease with increasing orbital separation \citep{deMink:2007SMC}, the efficiencies inferred from Algol systems may not directly translate to the BBH progenitors considered here. Nonetheless, the qualitative picture from \citet{Sen:2026}, that a range of intermediate, sub-unity mass-transfer efficiencies is required to produce MRR, is broadly consistent with the range explored in this work and in the other population synthesis studies mentioned above, representing a valuable observational counterpart.

Our analysis focuses exclusively on the BBH population, however MRR has been noted in the context of NS--BH binaries \citep{Sipior:2004}, though we note that \citet{Broekgaarden:2021imbI} find the occurrence rate to be less than $1\%$ of such systems. MRR has also been shown to be a generic consequence of binary evolution for neutron star--white dwarf binaries \citep{Tauris:2000,toonen18nov}. Similarly, MRR may play a role in shaping the NS--NS binary populations. A systematic study of MRR across all compact-object binary classes would be a valuable direction for future work. 

Our results are also complementary to a growing body of work examining how the SMT channel shapes the BBH mass and mass-ratio distributions. \citet{Gallegos-garcia:2021} compare the rapid population synthesis code \texttt{COSMIC} against \texttt{MESA} simulations; they find that common envelope dominates the BBH formation pathway in \texttt{COSMIC} while SMT dominates in their \texttt{MESA} runs and that depending on BH donor mass and mass-transfer and common-envelope physics \texttt{COSMIC} overproduces the BBH MRD by a factor of 2-35, underscoring that assumptions about mass-transfer stability alter not only the MRR fraction as we show here but also the overall MRD. In a similar fashion, \citet{Bavera:2021} combine rapid and detailed codes to estimate the BBH MRD while investigating the effects of mass-transfer and common-envelope efficiency on BBH observables. In particular, they find that the SMT channel preferentially produces BBH mergers with chirp masses of $\sim 15\,\Msun$ and leads to a MRD of 25 $\rm Gpc^{-3} yr^{-1}$. SMT has also been shown to reproduce the peak at $\sim 9 \, \rm M_\odot$ in the primary-mass distribution and that the low mass end in general can be explained by SMT systems alone \citep{vanson:2022smt}.  \citet{Dorozsmai:2024} used the rapid code \texttt{SeBa} to show that uncertainties in the first SMT phase have a large impact on which channel contributes to the MRD, but that the overall merger rate and primary-mass distribution are not affected greatly. This highlights that different binary physics can lead to comparable demographics and underscores that, even without MRR considerations, inferring progenitor properties from GW observables remains difficult. \citet{Dorozsmai:2024} also find that higher mass-transfer efficiency increases the population of BBH from higher primary-masses ($ \gtrsim 20\,\Msun$), consistent with our findings for $\beta$ in Section~\ref{sec:results-impact-uncertainties-compas}. Recently, \citet{Xu:2025} used detailed \texttt{MESA} models to study SMT, self-consistently modeling the chemical structure of the main-sequence accretor. They find that the more massive BH always descends from the second-born black hole in their models and that the mass-ratio distribution of merging BBHs peaks at 0.7 rather than unity, similar to our \texttt{COMPAS} models.

\subsection{COMPAS vs SEVN}
An important outcome of our work is the discrepancy between the \texttt{COMPAS} and \texttt{SEVN} results. These differences largely stem from the contrasting implementation of stellar evolution in each code. \texttt{COMPAS} utilizes the parameterized models for stellar and binary evolution from \cite{Hurley:2000,Hurley:2002rf} respectively, in which fitting formulae are employed to estimate stellar evolution. This is in contrast to the methodology employed by \texttt{SEVN}, which uses an updated set of tracks and interpolates between neighboring tracks \textit{on-the-fly} to evolve stars. 

While the fiducial models in this paper utilize different stellar tracks, it has been shown that discrepancies still arise between rapid and interpolated codes even when using identical tracks, as shown by \cite{Agrawal:2020MNRAS,Agrawal:2023}. Their work shows discrepancies arise in the core and remnant masses, lifetime, and radii with their interpolating code (\texttt{METISSE}) retaining finer details in reproducing the stellar tracks. 

Similarly, \cite{Romagnolo:2023} find discrepancies in the maximum radius--mass relationship between the \citet{Hurley:2000} (\texttt{SSE}) fitting formulae and 
\texttt{METISSE}. This work also reports differences in the total and component mass distributions between codes, becoming most pronounced for component masses of $\gtrsim 25 \Msun$, depending on the choice of tracks. The authors explain that this is also due to the fact that uncertainties increase with stellar masses, in particular for the regime $M_{\rm ZAMS}\gtrsim 50 \Msun$, where {\tt SSE}-based models require extrapolations of the \cite{Hurley:2000} fitting formulae.

Other explorations have been carried out, for instance by \citet{Belczynski:2022} who compared rapid population synthesis codes (\texttt{StarTrack} and \texttt{COMPAS}) to detailed evolution code \texttt{MESA}. In particular, they attempt to match the observational properties of binary system Melnick 34 finding a wide variety of remnant outcomes ranging from a close BBH system to a totally disrupted system, showcasing that evolutionary conclusions from codes must be interpreted with care.

Most directly relevant here, \citet{Guerrero:2026} carry out a systematic comparison of BBH formation across multiple population synthesis codes, isolating the impact of the stellar evolution treatment while holding the binary physics and initial conditions fixed as far as possible. 
They find that inconsistencies in the predicted BBH populations persist even under matched assumptions, reinforcing that a substantial part of the code-to-code scatter originates in the single-star treatment rather than in the binary-physics prescriptions or the sampling ranges.
This supports our interpretation that the qualitatively different MRR imprints in \texttt{COMPAS} and \texttt{SEVN} are driven primarily by how each code models stellar structure and its response to mass loss. A concrete example is the mapping from progenitor mass to final black hole mass at high metallicity, which is set largely by stellar winds and by whether a star collapses as a cool supergiant or is first stripped to a Wolf--Rayet star \citep[e.g.][]{Romagnolo:2026, Romagnolo:2026b}. Because \texttt{COMPAS} and \texttt{SEVN} adopt different wind implementations and different underlying tracks (Section~\ref{sec:methods}), this bifurcation alone can shift where in the mass distribution MRR systems appear, independently of the binary-evolution assumptions.

We note that the discrepancies arising from the stellar treatment in the fiducial models (e.g. Figure~\ref{fig:mass-distributions-MRR-vs-nonMRR}) is larger than those arising from single parameter variations (Figure~\ref{fig:compas-variations}). More plainly, no matter which \texttt{COMPAS} model is employed, there is always a regime in which MRR dominates in the mass distributions. While we did not explore single parameter variations in \texttt{SEVN}, the fiducial model contains no region where MRR is dominant in the mass distributions. 

Within \texttt{COMPAS}, the effects of MRR can lead to the \textit{massquerading} effect in the mass distributions most clearly shown in Figure~\ref{fig:Massquerading-in-LVK}. However, the effects of MRR in the \texttt{SEVN} results are hardly noticeable in this figure and are shown to be subdominant across the mass distributions (Figure~\ref{fig:mass-distributions-MRR-vs-nonMRR}).


We stress that these differences do not undermine our central, relative conclusion. The absolute merger rate and the detailed shape of the mass spectrum are known to be sensitive to stellar and binary modeling, and to the assumed star-formation history, at the level of orders of magnitude.
Indeed, a compilation of BBH merger rates across dozens of isolated binary-evolution studies shows that predictions span several orders of magnitude and that a substantial subset reproduces or underestimates the LVK-inferred rate, reflecting strong sensitivity to natal kicks, common-envelope evolution, mass transfer, angular-momentum loss, remnant formation, stellar winds, initial conditions, and star-formation history
\citep{Broekgaarden2026LYR, Broekgaarden2026CE}. Our finding that MRR contributes approximately one-third of the local merging BBH population is recovered independently in both codes despite these differences, which is interesting that is appears as the more robust of our results. A fully controlled one-to-one comparison, in which identical tracks, initial conditions, and binary-physics prescriptions are evolved through both frameworks so that only the single-star treatment differs, would allow these effects to be disentangled quantitatively and is a well-motivated direction for future work.

\subsection{Uncertainties in high-mass primary ZAMS systems}
Stellar evolution represents one of the key uncertainties in the formation of BBH systems. Following for instance \citet{Iorio:2023sevn}, the stellar tracks of which are used in our analysis with \texttt{SEVN}, at high metallicity and high initial masses stars do not significantly expand due to the ejection of their hydrogen-rich envelopes from strong winds (this is also highlighted by other recent studies such as \citealt{Romagnolo:2024maxBHmass,Hirschi:2025} and \citealt{Romagnolo:2026b}), with rotation potentially extending this phenomenon also down to Small Magellanic Cloud metallicities \citep{Boco:2025}. Non-expanding stars are not likely to initiate mass transfer events, and even if they do there is a limit on how close they can get from mass transfer \citep{Klencki:2026-smt}, which means that, i) such binaries are unlikely to contribute to the population of GW sources since their stellar components may evolve in isolation, and ii) the only available MRR for such systems is through PPISN-shrinking. However, PPISN can only happen for massive cores that are usually produced from the evolution of low-metallicity stars. This suggests that the high-M$_{\rm 1,ZAMS}$  population of PPISN MRR binaries may represent only a small sample of low-metallicity systems, with no contribution from the high-Z and high-M$_{\rm 1,ZAMS}$  binaries (see also \citealt{Romagnolo:2025}).
This is however only something that is clearly visible from our \texttt{SEVN} simulations, since codes that were derived from \cite{Hurley:2000,Hurley:2002rf} tend by design to expand considerably even after strong wind-driven mass loss \citep{Agrawal:2020MNRAS,Romagnolo:2023}.

\subsection{Mass accretion onto the first born black hole}
Throughout this work we assumed Eddington-limited mass accretion onto the first-born black hole from its stellar companion. This inherently limits how much mass the black hole can accrete, and therefore affects the rate of MRR. If super-Eddington black hole growth were allowed instead, the first-born black hole in some of our MRR systems could grow to become the more massive component once again, effectively reversing the mass ratio reversal. Note that \citet{Bavera:2021} show that if the BH accretion is not Eddington limited, then the contribution from SMT can drop by two orders of magnitude due to conservative mass transfer shrinking the binary orbit less efficiently.
Super-Eddington scenarios have already been invoked in the context of PISN-mass-gap black holes \citep{vanSon:2020,Woosley:2021xba,Briel:2023}, and sustained super-Eddington accretion has since been shown to be feasible \citep{ghodla:2023}, making this an interesting avenue to pursue in future work.

\section{Conclusions}
\label{sec:conclusions}
In this work, we have investigated \ac{MRR}, the evolutionary outcome in which the initially less massive star in a binary produces the more massive black hole, and assessed its imprint on the BBH population inferred by LVK. By analyzing population synthesis predictions from \texttt{COMPAS} and \texttt{SEVN} with merger rate modeling, we find that while MRR systems constitute a similar proportion of the local merger rate in both codes, they play a disproportionate role in shaping the high-mass and high-mass-ratio (approaching unity) regions of the BBH parameter space. Our results are summarized below:

\begin{enumerate}
    \item The local ($z\approx0.2$) MRR fraction is remarkably consistent between codes ($\sim34\%$ in \texttt{COMPAS}, $\sim35\%$ in \texttt{SEVN}), but this agreement breaks down across cosmic time: in \texttt{COMPAS} the MRR fraction declines from $\sim34\%$ at $z=0$ to $\sim21\%$ by $z\approx1.8$, whereas in \texttt{SEVN} it remains comparatively flat, rising mildly from $\sim35\%$ to $\sim37\%$ over the same range (Figure~\ref{fig:drdz_mrr} and Figure~\ref{fig:drdz-mrr-evolution}).
    \item The MRR fraction is plotted for the component and mass-ratio distributions in Figure~\ref{fig:mass-distributions-MRR-vs-nonMRR}, in which we find that MRR dominates the high-mass ($\mone \gtrsim 20 \Msun$, $\mtwo \gtrsim 12 \Msun$) and high-mass-ratio ($q>0.6$) regime in \texttt{COMPAS}. For \texttt{SEVN}, MRR peaks at low masses and high mass ratio ($q > 0.7$), but never dominates the distributions. The MRR mapping is further emphasized in Figure~\ref{fig:m1m2_2D} with a clear overabundance in the high-mass / high-mass-ratio regime for \texttt{COMPAS}, but only a slight peak in MRR fraction near the $q=1$ boundary in \texttt{SEVN}.
    \item We analyze 10 single parameter variations from the fiducial \texttt{COMPAS} model (Figure~\ref{fig:compas-variations}), finding that though there are differences in the overall MRR structure, these systems are found preferentially in the high-mass regime across all models, with $57$--$87\%$ of MRR systems lying above $15 \Msun$.
    \item MRR systems are mapped from their initial parameter space to their LVK defined black hole space, i.e. \mone being the larger BH, in Figure~\ref{fig:mrr_masses} (and to their ZAMS definition, \mone the initially more massive star in Figure~\ref{fig:ZAMS-BBH-properties-binary-M1M2}). We find that in \texttt{COMPAS}, MRR fractions are highest for  $M_{1,\rm ZAMS} \sim 50$--$ 90 \Msun$ , $M_{2,\rm ZAMS}\sim30$--$140 \Msun$,  $\log_{10}(Z/Z_\odot)\simeq-2.5$ to $-1$, and $q_{\rm ZAMS}\gtrsim0.5$. These get mapped to primary BHs with $20$--$40 \Msun$ and mass ratios of $q>0.6$. In \texttt{SEVN}, MRR fractions are highest for $M_{1,\rm ZAMS} \gtrsim  100 \Msun$ , $M_{2,\rm ZAMS}\gtrsim 80 \Msun$,  $\log_{10}(Z/Z_\odot)\simeq-4.0$ to $-0.4$, and $q_{\rm ZAMS}\gtrsim0.6$. However, for \texttt{SEVN} the mapping is dispersed across the BH mass space and no clear region has a dominating MRR fraction consistent with the findings in Figures~\ref{fig:mass-distributions-MRR-vs-nonMRR} and ~\ref{fig:m1m2_2D}.
    \item The primary evolutionary channel in which a binary becomes an MRR system is through the core-growth channel (Figure~\ref{fig:mrr_channels}), where the secondary core is grown larger than the primary star's through stable mass transfer. We find this channel accounts for $99\%$ of MRR systems contributing to the MRD at $z\sim 0.2$. We additionally identify three other MRR pathways that contribute $\lesssim1\%$ to the MRR BBH merger rate density at $z\sim0.2$. This includes:
    \begin{itemize}
        \item \texttt{SEVN}: Two subdominant channels are found, the first is PPISN-shrinking in which asymmetric pulsational mass loss leads to the primary star losing a substantial amount of its mass before collapse while the secondary retains a larger fraction of its envelope and becomes the larger BH. The last channel, asymmetric-CCSN, is attributed to uneven mass loss at the CCSN event, sometimes due to the secondary retaining its envelope.
        \item \texttt{COMPAS}: A variant of the core-growth channel, in which the secondary's core does not exceed the primary's through stable mass transfer alone; instead, PPISN mass loss pushes the primary to a slightly lower remnant mass than the secondary's, completing the reversal.
    \end{itemize}
    \item Finally, we find that the labels given by LVK, i.e. \mone and \mtwo referring to the more and less massive BH component do not map directly to the ZAMS definitions of donor and accretor. This \textit{massquerading} where the the secondary star “masquerades” as the primary in the observed BH mass distribution can lead to excess structure in the mass distributions which needs to be carefully considered when mapping between observed quantities and progenitor properties.
 
\end{enumerate}

We emphasize that \texttt{COMPAS} and \texttt{SEVN} give qualitatively different answers to the question of whether MRR reshapes the \emph{astrophysical} BBH population: in \texttt{COMPAS}, MRR dominates the high-mass and high-mass-ratio regime, whereas in \texttt{SEVN} it remains subdominant across the full mass distribution (Section~\ref{sec:results}, Figure~\ref{fig:mass-distributions-MRR-vs-nonMRR}). Despite this disagreement on where (or whether) MRR leaves a distinct imprint on the mass spectrum, both codes agree that MRR systems contribute a non-negligible $\sim 30\%$ of the local merger rate density (Figure~\ref{fig:drdz_mrr}), and that MRR preferentially populates the high-mass-ratio regime ($q\gtrsim0.6$). This $\sim30\%$ contribution to the MRD is the more robust of our two findings: it means that for roughly a third of the merging BBH population, the observed mass ordering does not necessarily trace the initial one, independent of whether a given model predicts MRR to dominate a specific region of parameter space. Taken together, our results demonstrate that MRR is not a rare outcome of stellar evolution, but a structurally important feature of isolated binary evolution. These systems must be carefully integrated into inference studies to ensure proper mapping between progenitor properties and observed quantities.

\subsection*{Data Availability}
For the \texttt{COMPAS} models the authors made use of the binary black hole simulations from \citep{Broekgaarden:2022imbII}, which are publicly available at  \citet[][]{Broekgaarden:2021-zenodo-BHBH}. The \texttt{SEVN} model was based on the simulations from \citet{Smith:2026}, using the same analysis pipeline we used an updated version of \texttt{SEVN (v2.13.0)} \citet{Iorio:2023sevn}. The \texttt{SEVN} data is publicly available along with necessary datasets to recreate all figures in \citet{smith_2026_20189849}. The observational data from LVK was obtained from the publicly available dataset at \citet{gwtc5:dataset}.

\subsection*{Software Acknowledgment}
This work made use of the following software packages: \texttt{astropy} \citep{astropy:2013,astropy:2018,astropy:2022,astropy_8408438}, \texttt{Jupyter} \citep{2007CSE.....9c..21P,kluyver2016jupyter}, \texttt{matplotlib} \citep{matplotlib}, \texttt{numpy} \citep{numpy}, \texttt{pandas} \citep{mckinney-proc-scipy-2010,pandas_19340003}, \texttt{python} \citep{python}, \texttt{SSPC} \citep{Hendriks:2023}, and \texttt{scipy} \citep{2020SciPy-NMeth,scipy_11255513}.

Software citation information aggregated using \texttt{\href{https://www.tomwagg.com/software-citation-station/}{The Software Citation Station}} \citep{software-citation-station-paper,software-citation-station-zenodo}.

\begin{acknowledgments}
TBS acknowledges support from the National Science Foundation Graduate Research Fellowship Program under Grant No. 1839285. 
FSB and SAL acknowledge support from NASA HPOSS grant 80NSSC25K7555 under award number 316592-00001.  FSB,  SAL, NJ, MS, and KAR acknowledge support from National Science Foundation 22-624 Astronomy and Astrophysics Research Grant under award number 2606407. 
AR acknowledges the support from the Polish National Science Center (NCN) grant Maestro (2018/30/A/ST9/00050). AR acknowledges financial support from the European Research Council for the ERC Consolidator grant DEMOBLACK, under contract no. 770017 and from the German Excellence Strategy via the Heidelberg Cluster of Excellence (EXC 2181 - 390900948) STRUCTURES. MK ackowledges support from the The University of California Leadership Excellence through Advanced Degrees (UC LEADS) program. 
\end{acknowledgments}

\bibliography{mrr}
\bibliographystyle{aasjournal}

\appendix
\twocolumngrid

\section{Mass ratio criterion}
\label{app:mrr-threshold}

In this work we have used the simple criteria that the mass of the black hole descending from the initially less massive star (B) needs to be larger than the initially more massive star's black hole (A), i.e. M$_{\rm B}$ > M$_{\rm A}$. However, the criteria for MRR can be shifted to be more conservative, for instance it can be set to M$_{\rm B}$ > 1.05 M$_{\rm A}$. However, this can unfairly impact massive stars where the difference can be multiple \Msun. It is also possible to present the criteria as a minimum mass difference between the two, e.g. M$_{\rm B}$ > M$_{\rm A}$ + 1 M$_{\odot}$. This on the other hand can be more difficult for smaller mass systems to overcome. Thus, we simply use M$_{\rm B}$ > M$_{\rm A}$ in this work, but highlight one possible alternative in this appendix. 

Comparing the fractional contribution from MRR systems in the total MRD at $z \approx $ 0.2, we find that in our current approach \texttt{COMPAS} sits at $33\%$, while \texttt{SEVN} at $32\%$. If we implement a threshold of M$_{\rm B}$ > 1.05 $\times$ M$_{\rm A}$ the fractional contributions are: \texttt{COMPAS} - $29\%$ and \texttt{SEVN} - $27\%$. These are relatively negligible differences, and this level persists across the full redshift evolution. 

While, these are small differences in the overall MRD, the differential MRD tells a slightly different story. Figure~\ref{fig:mass-distributions-MRR-vs-nonMRR} shows the case with no threshold and we include in Figure~\ref{fig:dRdX-mrr-criteria} the case for M$_{\rm B}$ > 1.05 M$_{\rm A}$. For \texttt{COMPAS} the differences are overall minimal, there exists a peak in the primary- and secondary-mass distributions at $\sim 25 \, \Msun$, the non-MRR contribution dominates at the very end of the secondary distribution ($\sim 36-40 \Msun$), and the high-mass-ratio regime falls off for MRR systems (as expected due to the cut being placed on mass ratio itself).  Similarly, the results from \texttt{SEVN} remain consistent with MRR never dominating, however the MRR contribution drops off much steeper when implementing a conservative cut. Overall, the results are largely consistent including a conservative cut, the total MRD remains nearly the same, and the main behavior in dR/dX remains consistent.

\begin{figure*}
    \centering
    \includegraphics[width=\linewidth]{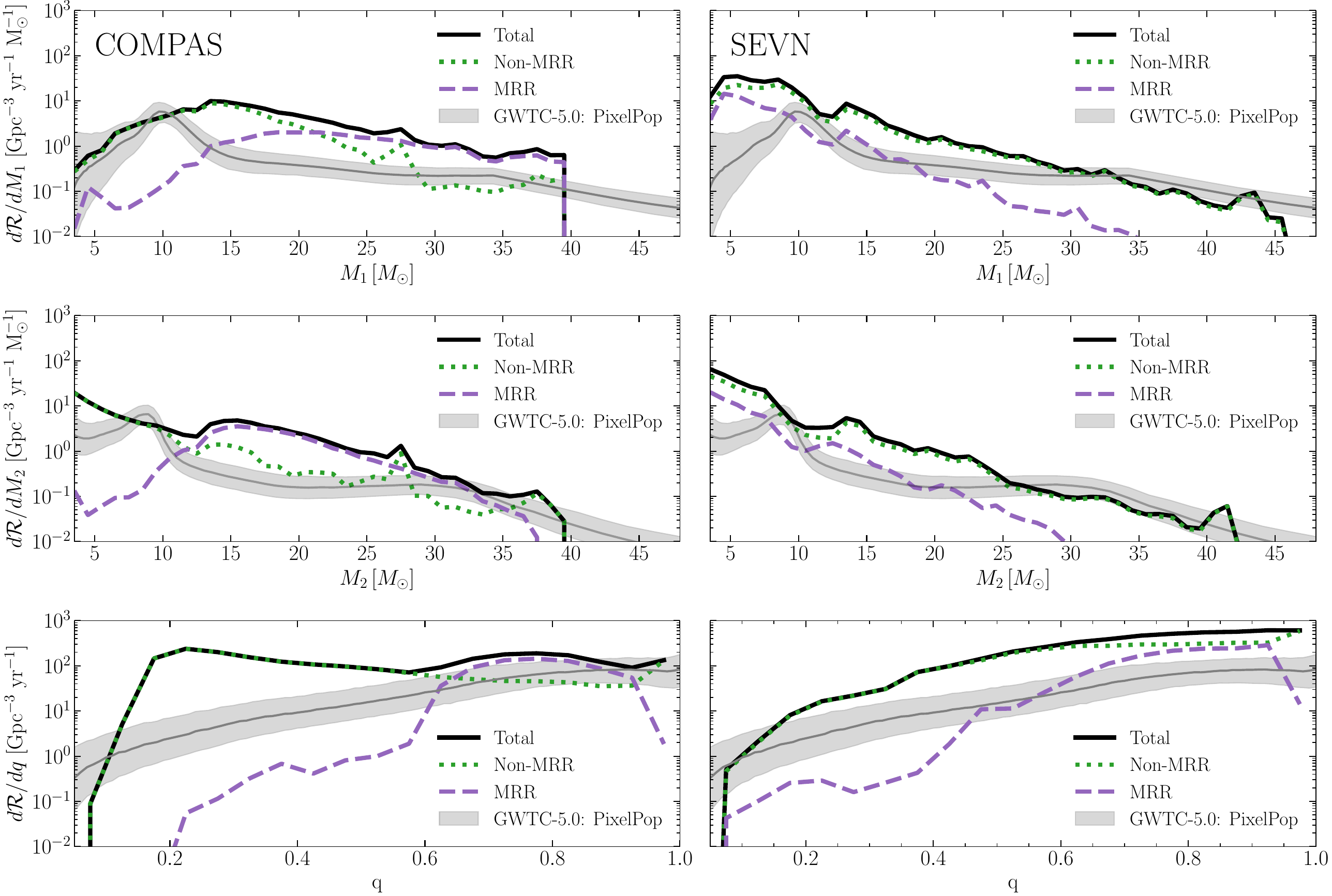}
    \caption{The same as Figure~\ref{fig:mass-distributions-MRR-vs-nonMRR}, but with an imposed threshold that the secondary star's remnant needs to be at least $5\%$ more massive than the primary star's remnant (M$_{\rm B}$ > 1.05 $\times$ M$_{\rm A}$). For \texttt{COMPAS} (left-column) the main difference in the primary- and secondary-mass distributions are that the non-MRR contribution increases with a peak between $25$--$30$ \Msun and a dip at mass ratios near unity (since this is where the cut-off lives. In \texttt{SEVN} the same behavior is seen in the mass-ratio distribution, and in the primary- and secondary-mass distributions we note that there is a steeper drop off from the MRR contribution tending towards higher masses.}
    \label{fig:dRdX-mrr-criteria}
\end{figure*}

\section{Variations in the chirp mass distribution}
\label{app:chirp-mass}

We present an analysis on single-parameter variations in \texttt{COMPAS} for the chirp-mass distribution in Figure~\ref{fig:compas-variations-chirp}. The majority of the models exhibit a pronounced peak at $\mathcal{M}_c \simeq 5 \Msun$, which is notably more robust than the peak of the BH primary-mass distribution in Section~\ref{sec:results}. This low-chirp-mass peak is dominated by non-MRR systems and remains largely unchanged across all explored prescriptions considered here.

\begin{figure*}[h!]
    \centering
    \includegraphics[width=0.9\linewidth]{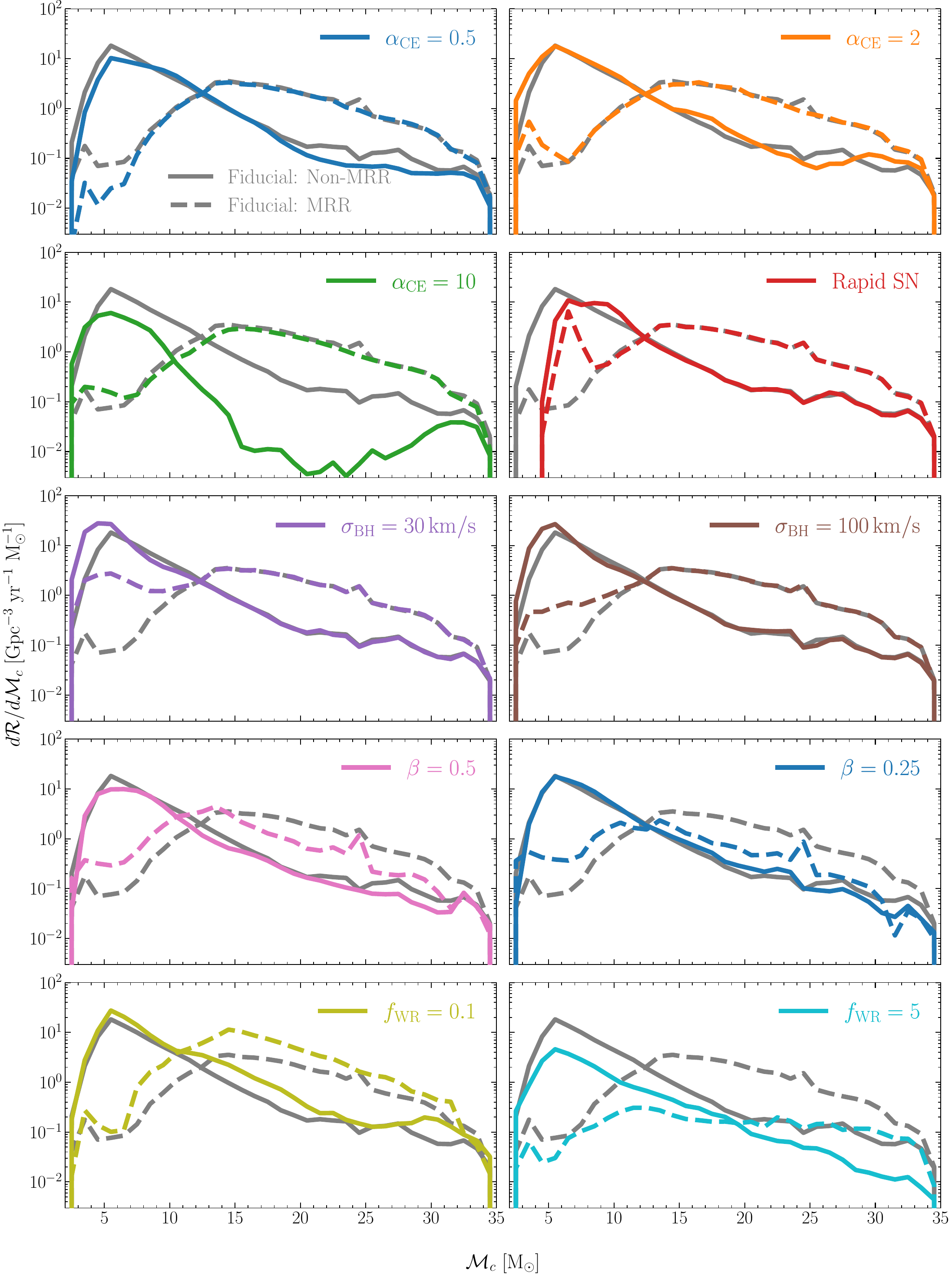}
    \caption{The chirp-mass distribution is shown for variations in binary parameters relative to the fiducial model, following \citet{Broekgaarden:2022imbII}. Each panel compares the MRR and non-MRR contributions for the fiducial model and a single parameter variation. While individual prescriptions modify the overall normalization and relative MRR contribution, the low-mass peak, dominated by non-MRR systems, remains robust across each variation.}
    \label{fig:compas-variations-chirp}
\end{figure*}

The largest variations in the chirp-mass spectrum arise from changes to the mass-transfer efficiency $\beta$ and the Wolf-Rayet wind strength $f_{\rm WR}$. Increasing $\beta$ enhances the relative contribution of MRR systems at lower chirp masses while suppressing their contribution at higher chirp masses, with the non-MRR population remaining comparatively stable. 
Variations in $f_{\rm WR}$, particularly for stronger winds, flatten the MRR contribution across the chirp-mass range and reduce the non-MRR rate, producing the largest overall change among the tested parameters. In contrast, variations in the common-envelope efficiency $\alpha_{\rm CE}$, supernova engine prescription, and the BH natal kick dispersion $\sigma_{\rm BH}$ primarily affect the low-mass end of the chirp-mass distribution, akin to the primary-mass distribution, with limited impact on the relative MRR contribution at higher chirp masses.

While much of our discussion in Section~\ref{sec:discussion} focused on the component-mass and mass-ratio distributions, chirp mass provides a complementary and observationally robust diagnostic. We find that the low-chirp-mass peak, which is dominated by non-MRR systems, is stable across a wide range of binary-evolution prescriptions. In contrast, MRR systems preferentially contribute at higher chirp masses, with their relative importance varying modestly under changes to mass transfer efficiency and Wolf-Rayet wind strength. These results demonstrate that while the overall normalization and detailed shape of the chirp-mass distribution can vary, the qualitative separation between MRR- and non-MRR-dominated regimes is robust to reasonable changes in binary-evolution physics. This robustness strengthens the case for using chirp mass, in conjunction with component masses and mass ratio, to assess the potential contribution of MRR in LVK observations.

\section{MRR channels - formation efficiency yields}
\label{app:mrrchannels-formation-eff}

We showed in Section~\ref{subsec:MRR-channels} that MRR can arise from several distinct evolutionary channels. In particular, that core-growth is the dominant channel in both \texttt{COMPAS} and \texttt{SEVN} when the distributions are weighted by the BBH merger rate at $z \approx 0.2$. We discuss in this appendix the initial parameter space when weighted instead by the formation efficiency (Equation~\ref{eq:Nform}). 

In \texttt{COMPAS}, there are minimal differences with the core-growth contribution dropping to $98\%$ of the total MRR contribution. However, the PPISN-assisted core-growth channel does dominate in small regions of initial parameter space in particular at high primary masses of $M_{1, \rm ZAMS} \gtrsim 120 \Msun$, secondary masses of $110$--$125 \Msun$, and low mass ratios (Figure~\ref{fig:mrr_mass_channel_form_eff}). 

In \texttt{SEVN} we find that despite both PPISN-shrinking and asymmetric-CCSN almost never being the dominant MRR GW channel in \texttt{SEVN} ($0.5\%$ and $1.5\%$ respectively), their formation yields are fundamentally more efficient ($4.5\%$ and $3.5\%$ respectively) than what can be observed due to their formation likelihood. 

The contribution from the PPISN-shrinking channel has the most noticeable difference between the two weighting schemes. For the primarymass distribution (left most column in Figure~\ref{fig:mrr_mass_channel_form_eff}), the core-growth and asymmetric-CCSN channels are largely similar to that of the MRD weighted distributions. On the other hand, while the PPISN-shrinking channel contributed minimally to the MRD weighted distribution, it has a much stronger contribution to the overall formation-efficiency (which is marginalized over metallicity). We note that due to the metallicity evolution of the universe, the contribution of the PPISN-shrinking channel may vary with cosmic time, i.e. for the MRD at earlier times we expect the fraction to increase. 

The effects on the secondary-mass and mass-ratio distribution are more subtle with PPISN-shrinking dominating for M$_{\rm 2, ZAMS} \gtrsim$ 100 $\Msun$ and $q \lesssim0.7$.

Lastly, the metallicity distribution shows that the relevant metallicities for the MRD at $z \approx 0.2$ peaks at higher metallicities (Figure~\ref{fig:mrr_mass_channel_MRD}) with asymmetric-CCSN and PPISN-shrinking remaining subdominant over the full metallicity range. On the other hand, the formation-efficiency weighted distribution favors low metallicity. This is a reflection of the well-known result that formation-efficiency is highest for low metallicities and decreases as metallicity approaches solar metallicity
\citep[e.g.][]{Belczynski2010,mapelli2010,Stevenson:2017tfq, Chruslinska:2018hrb, Giacobbo:2018etu, Neijssel2019, Spera2019, Broekgaarden:2022imbII, Schiebelbein-Zwack:2024roj, Smith:2026}. 

To summarize, the formation-efficiency weighted distributions have larger contributions from ppisn-assisted core-growth in \texttt{COMPAS} and asymmetric-CCSN and PPISN-shrinking in \texttt{SEVN}. While these differences mostly arise from the marginalization over metallicity/redshift, they can lead to these channels dominating in particular regions of initial parameter space.

\begin{figure*}
    \centering
    \includegraphics[width=\linewidth]{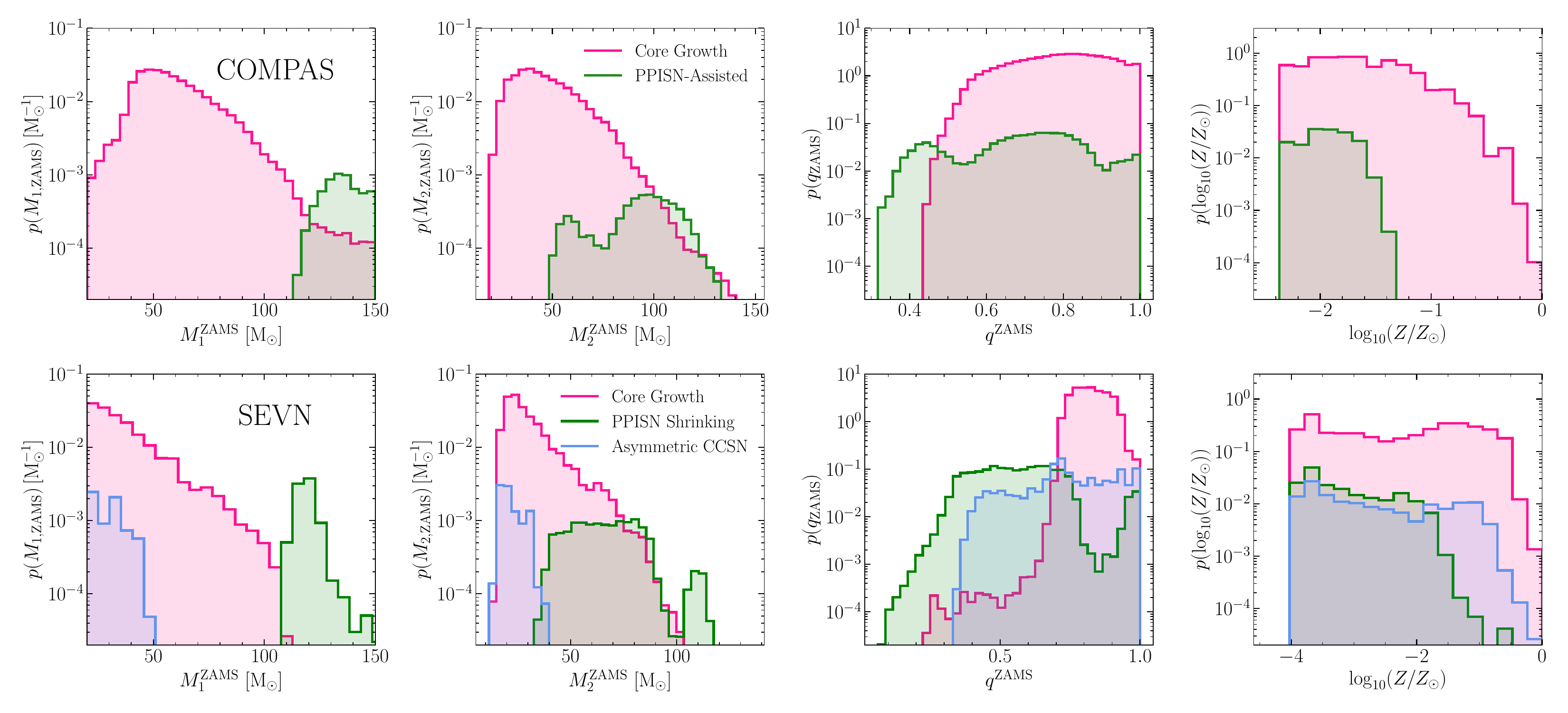}
    \caption{Same as Figure~\ref{fig:mrr_mass_channel_MRD}, but instead weighted by the formation efficiency.}
    \label{fig:mrr_mass_channel_form_eff}
\end{figure*}

\section{Mass ratio reversal dependence on orbital separation}
\label{app:orbital-sep}

The orbital separation is sampled in \texttt{COMPAS} from [0.1, 1000] AU and from [0.07, 670] in \texttt{SEVN}. Figures~\ref{fig:orb-sep-form-eff} and \ref{fig:orb-sep-yield} show contributions from MRR and non-MRR systems weighted by the formation efficiency and merger rate yield respectively for both \texttt{COMPAS} (left) and \texttt{SEVN} (right). Included for both codes is a deconstruction of the MRR contribution into constituent channels. 

In both \texttt{COMPAS} and \texttt{SEVN}, the majority of merged BBH systems and the overall MRR fraction have a preference for short semimajor axes. The results by \citet{schneider2015evolution} show that orbital separation is a good proxy for determining which mass transfer case a system underwent. For instance, case A typically dominates below $10$--$100$ AU, above this case B takes over up to several thousand solar radii followed by case C with a strong dependence on stellar mass. In these approximations, the majority of MRR systems undergo case A mass transfer. It would be interesting for future work to quantify the dependence and fraction of MRR on various mass transfer cases. 

In \texttt{SEVN}, the PPISN-shrink channel peaks at slightly higher orbital separations, as slightly wider separations are less likely to disrupt the binary. The total fraction of MRR systems originating from the PPISN-shrink channel contribute a negligible fraction to the merger rate density. Asymmetric-CCSN follows a similar pattern to core-growth, but remains subdominant throughout both formation and MRD weighting. Similarly, in \texttt{COMPAS} the overall contribution from the PPISN-assisted core-growth channel is small compared to the dominant core-growth channel.

\begin{figure*}
    \centering
    \includegraphics[width=\linewidth]{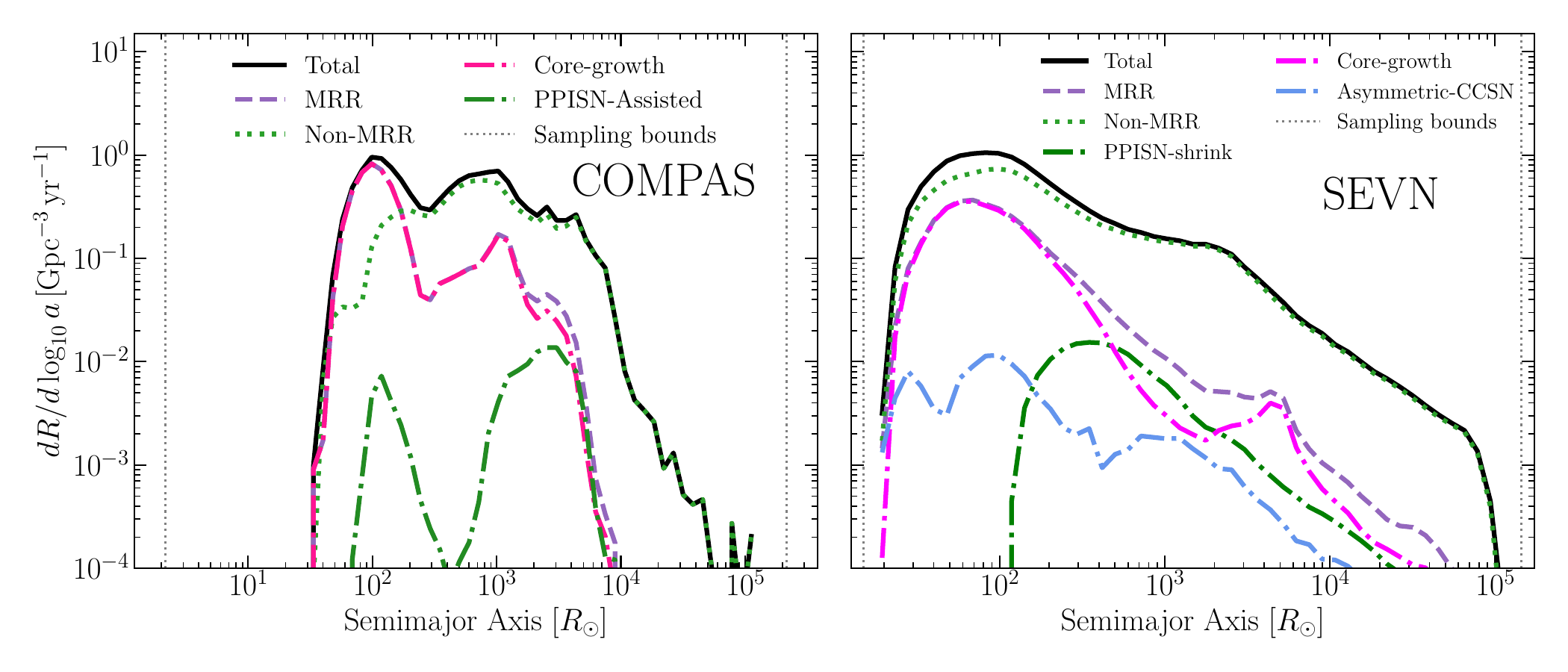}
    \caption{MRR and non-MRR contributions as a function of initial progenitor semimajor axis for \texttt{COMPAS} (left) and \texttt{SEVN} (right). Included is the evolutionary channel decomposition of the MRR contribution. This distribution is weighted by the formation efficiency (Equation~\ref{eq:Nform}).}
    \label{fig:orb-sep-form-eff}
\end{figure*}

\begin{figure*}
    \centering
    \includegraphics[width=\linewidth]{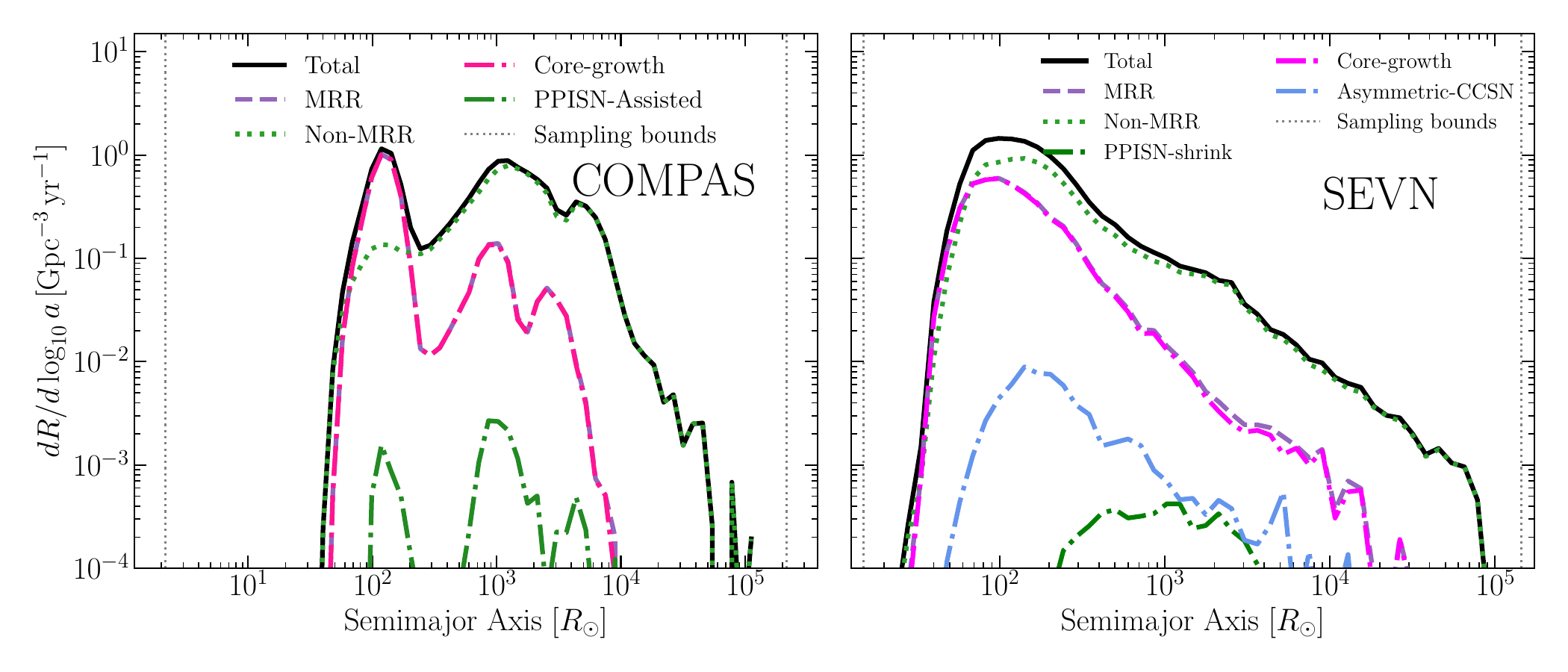}
    \caption{Same as Figure~\ref{fig:orb-sep-form-eff}, but weighting by the contribution to the MRD at $z\approx0.2$.}
    \label{fig:orb-sep-yield}
\end{figure*}

\section{Do variations alter the massquerading effect?}
\label{app:massquerading-variations}

We highlight in Figure~\ref{fig:compas-variations-massquerade} the effects of single parameter variations of the fiducial \texttt{COMPAS} model for the primarymass distribution. The results show that for each variation the \textit{massquerading} effect is strongest in the high-mass regime, typically around M $\gtrsim 20 \Msun$. This highlights that this effect is not dependent on any single stellar/binary parameter, but is a robust outcome of stellar evolution found in \texttt{COMPAS}. If MRR plays a large role in redefining the primary and secondary mass according to LVK definition, then the component masses will contain a mixture of donor and accretor remnants.

\begin{figure*}
    \centering
    \includegraphics[width=\linewidth]{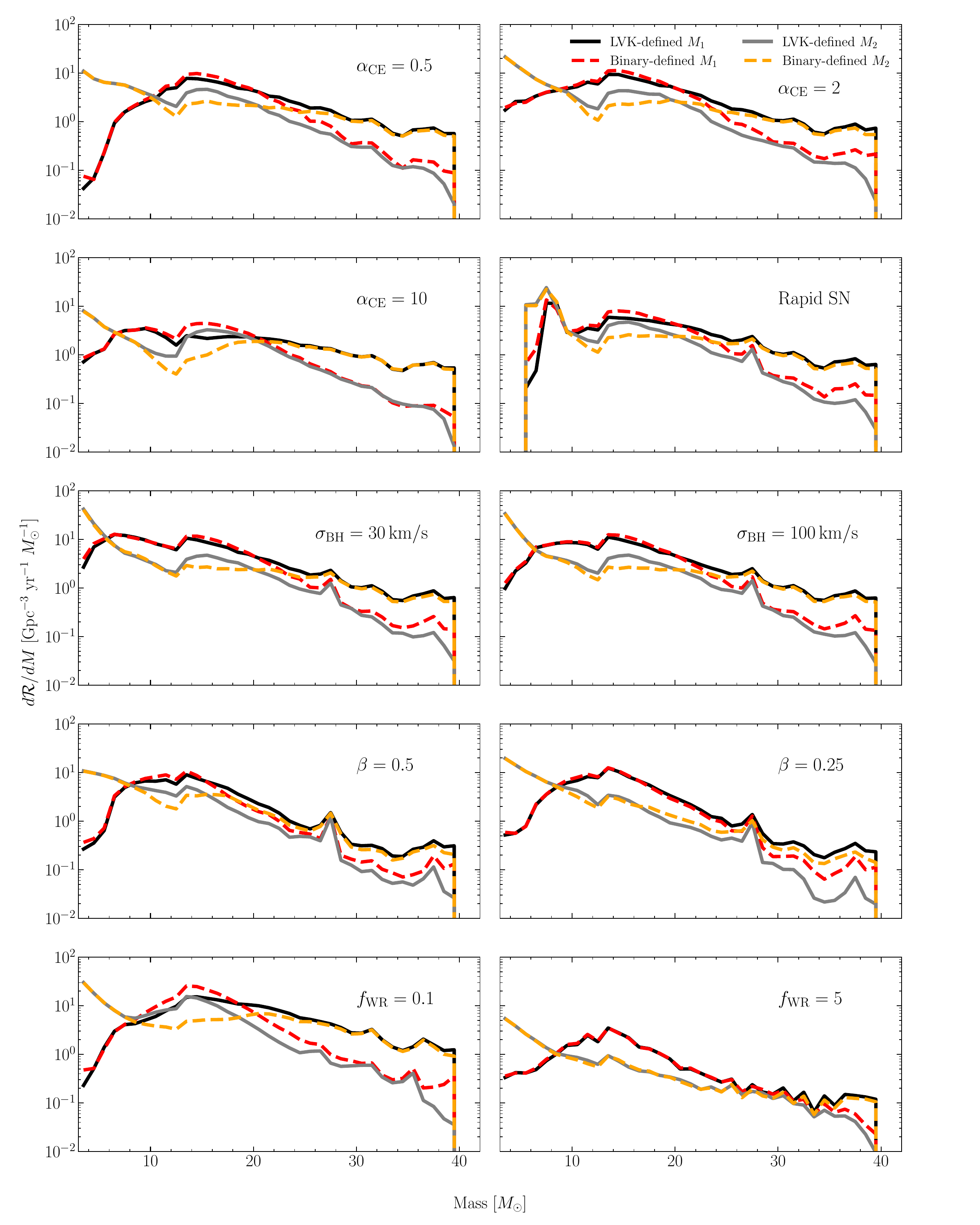}
    \caption{Variations of the massquerading effect based on single-parameter variations in \texttt{COMPAS}. The effect occurs mostly for $\gtrsim 20 \Msun$ as this is where MRR constitutes a dominant fraction of systems. See Figure~\ref{fig:Massquerading-in-LVK} for massquerading in the fiducial model.}
    \label{fig:compas-variations-massquerade}
\end{figure*}

\section{initial star properties mapping to binary defined BBH properties}
\label{sec-appendix-ZAMS-BBH-properties-binary-M1M2}
We show in Section~\ref{subsec:mrr-origins} how the initial parameter space is populated with MRR systems and the mapping, across metallicities, to the "LVK" defined mass space, i.e. \mone is the more massive BH and \mtwo the less massive. In this appendix, we also include the mapping between the initial parameters and the "binary" defined BHs (Figure~\ref{fig:ZAMS-BBH-properties-binary-M1M2}), i.e. initially more massive star is the primary.

Interestingly, \texttt{COMPAS} remains fairly consistent in the overall 2D distributions, while for \texttt{SEVN} a region of MRR abundance becomes clear in the low-metallicity and low-mass regime of the binary-defined primary BH and high-mass binary-defined secondary BH spaces. These systems are a small subset of the high-mass contribution and are still overcome by the non-MRR systems in the merger rate distributions (Figure~\ref{fig:mass-distributions-MRR-vs-nonMRR}).  

\begin{figure*}[h!]
    \centering
    \includegraphics[width=0.9\linewidth]{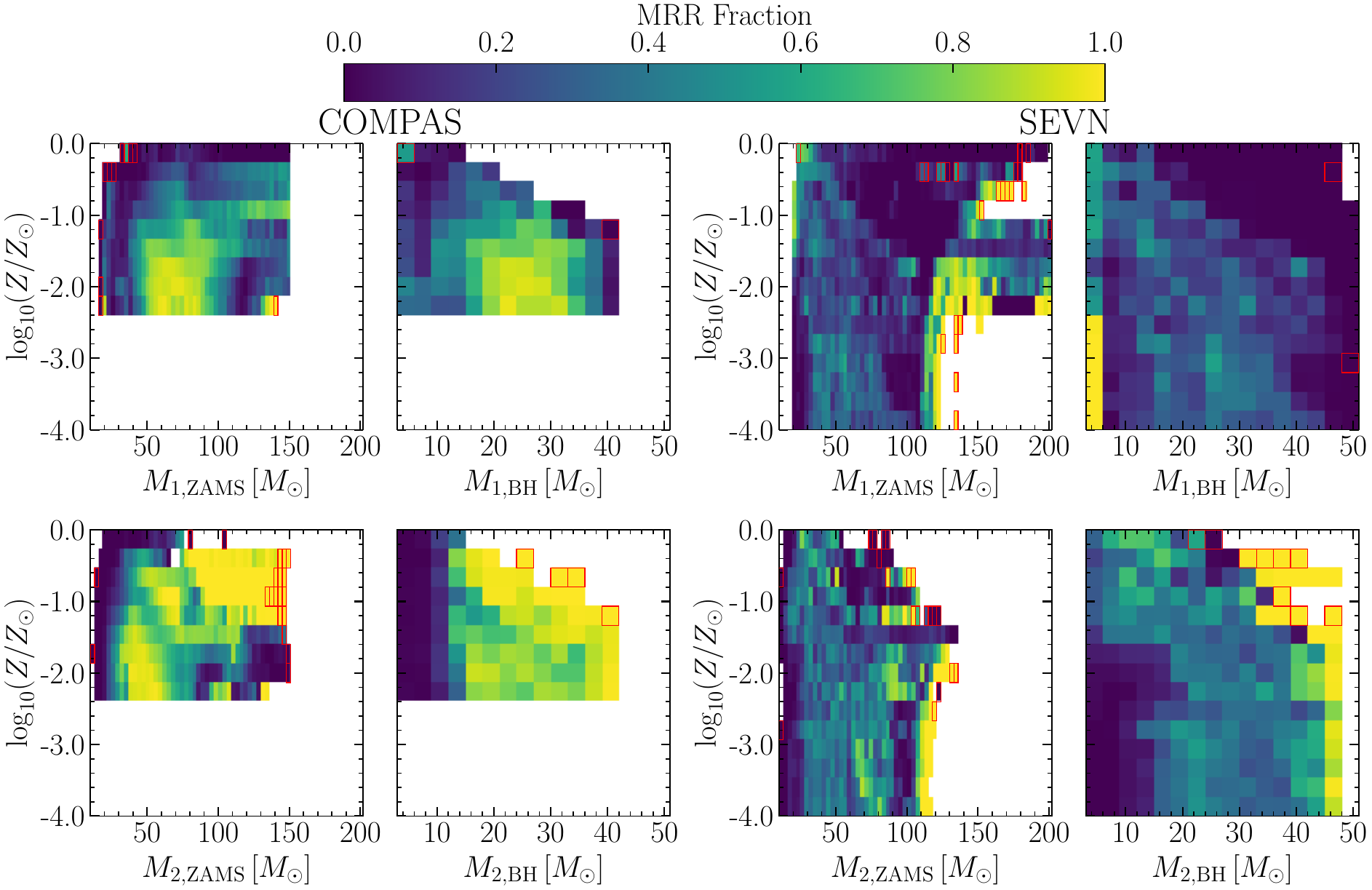}
    \caption{Same as Figure~\ref{fig:mrr_masses}, but for the remnant masses defined with \mone the most massive star at ZAMS and \mtwo the less massive. For \texttt{COMPAS}, the MRR fractions do not shift by a large amount. However, for \texttt{SEVN} an abundance of MRR systems becomes clear at low masses for \mone $\lesssim 10 \Msun$ and $Z\lesssim -2.5$, additionally MRR becomes dominant for $\mtwo \gtrsim 30 \Msun$.}
    \label{fig:ZAMS-BBH-properties-binary-M1M2}
\end{figure*}

\end{document}